 \documentclass[final,3p,times,twocolumn]{elsarticle}
\usepackage{epsf}
\usepackage{epsfig}
\usepackage{mathrsfs}
\usepackage{amsmath, amssymb}
\usepackage{subfigure}
\usepackage{color}
\usepackage{caption}
\usepackage{graphics} % for pdf, bitmapped graphics files
\usepackage{epsfig} % for postscript graphics files
\usepackage{times} % assumes new font selection scheme installed
\usepackage{amsmath} % assumes amsmath package installed
\usepackage{amssymb}  % assumes amsmath package installed
\usepackage{mathrsfs}
\usepackage{array}
\usepackage{algorithm}
\usepackage{algorithmic}

\usepackage{url}
\usepackage{array}
\usepackage{graphics} % for pdf, bitmapped graphics files
\usepackage{epsfig} % for postscript graphics files
\usepackage{times} % assumes new font selection scheme installed
\usepackage{amsmath} % assumes amsmath package installed
\usepackage{amssymb}  % assumes amsmath package installed
\usepackage{mathrsfs}
\usepackage{array}
\usepackage[english]{babel}
\usepackage{times}
\usepackage{psfrag}
\usepackage{dsfont}
\usepackage{algorithmic}
\usepackage{algorithm}
\usepackage{setspace}
\usepackage{amssymb}
\usepackage{amsmath}
\usepackage{graphicx}
\usepackage{psfrag}

\def\x{{\bf x}}

\usepackage{amssymb}
\journal{Computer Communications}

\begin{document}

\begin{frontmatter}



\title{Dynamic Frequency Management in 802.11-based\\ Multi-Radio Wireless Networks}

\author[label1]{George Athanasiou}
\address[label1]{Automatic Control Lab, School of Electrical Engineering, \\KTH Royal Institue of Technology, Sweden. E-mail: \textit{georgioa@kth.se}}

\author[label2]{Leandros Tassiulas}
\address[label2]{Department of Computers and Communications Engineering, \\University of Thessaly, Greece. E-mail: \textit{leandros@uth.gr}}

\address{}

\begin{abstract}
Efficient channel selection is essential in 802.11 mesh deployments, for minimizing contention and interference among co-channel devices and thereby supporting a plurality of QoS-sensitive applications. A few protocols have been proposed for frequency allocation in such networks, however they do not address the problem {\em end-to-end}.
In this paper, we present a general formulation of the channel selection problem taking into account the performance of both mesh-access and mesh-backhaul. Moreover, we propose ARACHNE, a routing-aware channel selection protocol for wireless mesh networks.
ARACHNE  is distributed in nature, and motivated by our  measurements on a wireless testbed.
The main novelty of our protocol comes from adopting a metric that captures the end-to-end link loads across different routes in the network.
ARACHNE  prioritizes the assignment of low-interference channels to links that
(a) need to serve high-load aggregate traffic and/or
(b) already suffer significant levels of contention and interference.
Our protocol takes into account the number of  potential interfaces (radios) per device, and allocates these interfaces in a manner that  efficiently utilizes  the available channel capacity.
We evaluate ARACHNE through extensive, trace-driven simulations and we show that approaches the optimal channel selection. We observe that our protocol improves the total network throughput, as compared to three other representative channel allocation approaches in literature.

\end{abstract}


\end{frontmatter}

% \linenumbers

%\section{}
%\label{}
\section{Introduction}
\label{sec:introduction}

Wireless mesh networking has been touted as the new technology that can support ubiquitous end-to-end connectivity.
In wireless mesh networks, information has to be routed via multiple wireless hops before it can reach the destination \cite{microsoftmesh, meraki}.
A critical requirement for the efficient routing of packets is the identification and use of interference-limited wireless links.
Therefore, intermediate mesh hops along a route need to operate in frequencies, where contention and interference are as low as possible, especially in highly-dense mesh deployments.
%In this paper, we
We ask the question:
{\em How can we allocate frequencies in a mesh network, in order to maximize the total network throughput, in a distributed manner?}

\begin{figure}
\centering
\includegraphics[width=2.5in]{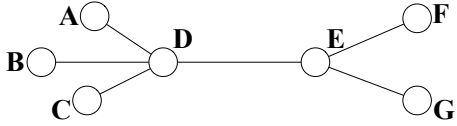}
\caption{ Load-aware channel selection.}
\label{example}
\end{figure}

In order to efficiently allocate the set of available channels to nodes, the load at each individual link needs to be taken into account.
Here, the load at the mesh access and backhaul levels could be represented in various ways \cite{Athanasiou,  gibbsInfocom}, potentially involving the number of neighbor transmitters, the traffic demand, the amount of traffic flowing through each node, etc., as we discuss later.
Previous studies on frequency selection, however, do not consider the {\em end-to-end load distribution} across entire routes;  they consider the sub-problems of either the access level, between clients and access points (APs) \cite{gibbsInfocom, rozner}, or the backbone portion of the network \cite{maxchop, hyacinth, mobicom05}.
We argue that frequency selection algorithms %, in order for frequency selection to provide near-optimal network capacity, it
 should prioritize the assignment of low-interference channels at highly-loaded mesh links, both at the access and the backhaul levels.
As a simple example, consider the connectivity graph of Fig. \ref{example}, where nodes $A$, $B$ and $C$ generate equal amounts of traffic towards node $E$, while the same channel is initially used by all links.

%
%\begin{figure}[h]
%\centering
%\includegraphics[width=1.3in]{figures/example.eps}
%\caption{\small Load-aware channel selection.}
%\label{example}
%\end{figure}
%
In this scenario, the $DE$ link facilitates the aggregated traffic, destined to node $E$.
Thus, $DE$ should be assigned a frequency such that the aggregate traffic towards $E$ is efficiently forwarded, i.e., the bottleneck situation with $DE$, due to potentially low SINR or high contention levels, is avoided to the extent possible.
Note here that the frequency selection outcome will likely affect the decision of load-aware routing protocols, such as RM-AODV (Radio Metric Ad Hoc On-Demand Distance Vector Routing) \cite{80211s}.
Hence, both the frequency selection and load-aware routing functionalities are inter-dependent and must be considered in conjunction.

In this paper we present a general formulation of the channel allocation problem, taking into account the important parameters (i.e., interference level, packet dropping and transmission rate) that affect the end-to-end throughput in the network. Moreover, we propose ARACHNE, a load and routing aware channel selection protocol for wireless mesh networks.
ARACHNE performs end-to-end channel selection along a route, by adopting a variation of a load characterization metric \cite{Athanasiou}.
Given that the load-aware routing choices are affected by the frequency selection policy, ARACHNE combines frequency selection and route selection under the same unified framework and approaches the maximum throughput in the network.
To the best of our knowledge, this is one of the first works that present a frequency selection protocol across entire routes, involving both the access and the backhaul levels in conjunction with the load-aware routing of information between end-hosts. We evaluate ARACHNE through trace-driven simulations, using the OPNET \cite{opnet} simulation platform. We observe that ARACHNE outperforms other channel allocation mechanisms for mesh networks in terms of overall network throughput, average delay and dropped data.

The rest of this paper is structured as follows.
%Section \ref{sec:related} provides the relevant previous work on channel allocation for mesh networks.
In section \ref{sec:prelimexp} we present our preliminary experiments and discuss the previous approaches related to our work, which motivate our problem formulation and the design of ARACHNE.
Section \ref{sec:theory} presents the modelling of channel allocation in wireless mesh networks and describes the airtime metric that we employ, based on which ARACHNE discovers the channel with the maximum long-term throughput.
In section \ref{sec:protocol}, we describe in detail the design of ARACHNE for both the access and the backhaul levels.
We evaluate our protocol through simulations, in section \ref{sec:evaluation}.
Finally, section \ref{sec:conclusion} concludes this paper.

%\section{Motivating the End-To-End Load-Aware Channel Allocation Policy}
\section{Motivating our Channel Allocation Policy}
\label{sec:prelimexp}

In this section, we describe a set of preliminary experiments on our wireless testbed, which motivate the design of ARACHNE.
%These measurements help realize that a channel allocation protocol for mesh networks should take into account the end-to-end routing behavior, starting with the most loaded point-to-point mesh links, and not only considering the interference experienced at individual links.
We first provide a description of our testbed platform, and subsequently we discuss our experiments and their interpretations.
We also discuss relevant previous work.

%
%\begin{figure}[t]
%\begin{center}
%\includegraphics[width=3.4cm,angle=270]{figures/prelim.ps}
%\caption{\small Taking into account the traffic load (A2) outperforms the interference-only aware channel policy (A1), especially at high traffic loads. }
%\label{fig:prelim}
%\end{center}\vspace{-0.3in}
%\end{figure}
%
%\begin{figure}[t]
%\begin{center}
%\includegraphics[width=5cm]{figures/ucr.eps}
%\caption{\small Deployment of our testbed.} % Node locations are represented by dots along with their IDs. }
%\label{fig:ucr}
%\end{center}\vspace{-0.3in}
%\end{figure}
\begin{figure}
\centering
\includegraphics[width=3.2in]{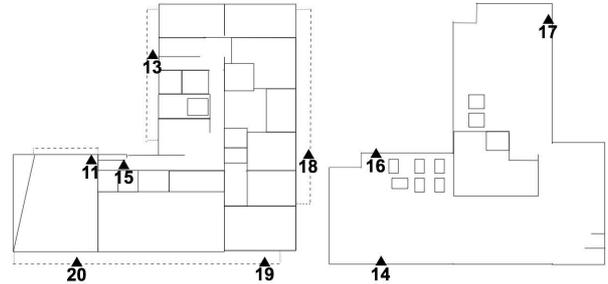}
\caption{The testbed deployment in the 4th (left) and the 5th (right) floor of our building.} \label{fig:ucr}
\end{figure}

Our wireless testbed deployment (Fig. \ref{fig:ucr}) %is deployed in the 3rd floor of a campus building
%.
%It
consists of 9 nodes that are based on the ORBIT hardware
configuration, and run a Debian Linux distribution
with kernel v2.6 over NFS. Each node is equipped with
1 GHz CPU, 512 Mbytes of memory, and a WN-CM9
wireless card, which carries the AR5212 Atheros chipset.
%42 Soekris net4826 nodes, %which run a Debian Linux distribution with kernel v2.6.
%each
%Each node is
%equipped with two EMP-8602 6G miniPCI 802.11a/b/g WiFi cards.
We set the cards to 802.11g mode and we use channels 1, 6 and 11.
%an \textit{EMP-8602 6G} with Atheros chipset  and
%an {\em Intel-2915}.
%Our measurements take place late at night, in order to avoid contention and interference from co-located WLANs; we set the ad-hoc 802.11g mode of operation on all interfaces and we use only the 3 non-overlapping 802.11g channels, 1, 6 and 11.
%Note that we measure the levels of {\em external} interference for all channels, and we conduct our experiments on the channels with negligible or zero {\em external} interference.
For the purposes of these preliminary experiments we consider fixed routes between source and destination.
%; for this, we manually insert these routes into the routing tables of nodes.
%We also make sure that some links in the network serve multiple routes.
We inject UDP traffic with various constant bit rates (CBR) and with packet size equal to 1500 bytes.
%The transmission power at all interfaces is 15 dBm.
%We initiate traffic between certain source-destination pairs, in parallel.
For each end-to-end traffic session we set different application data rates;
we utilize the {\em iperf} measurement tool.

We provide a representative experiment in what
follows. %\footnote{We perform experiments with 34 different pairs of routes and we observe the same behavior.}.
Consider the following two simultaneously active routes (see Fig. \ref{fig:ucr}):
{{\bf (a) 16}$\rightarrow$15$\rightarrow$20$\rightarrow${\bf 19}}, %with maximum data rate  10 Mbps,
and {{\bf (b) 13}$\rightarrow$15$\rightarrow$20$\rightarrow${\bf 14}.
%, with maximum data rate 1 Mbps.
These routes have one link in common, 15$\rightarrow$20, while all links are of similar quality in terms of achievable throughput in isolation.
We apply two different channel selection policies;
 for both policies we make sure  %(manually)
that connectivity is maintained between the end-hosts, {\em i.e., two consecutive relays use the same channel on one of their interfaces.}
We first consider the channel  policy, $A1$, which assigns channels according to the interference experienced (allocates the channel with the minimum aggregate interference, observed through RSSI measurements).
Subsequently we assign channels to nodes, additionally taking into account the link loads, in terms of both link quality and aggregate traffic service; we call this policy $A2$.
In other words, we manually prioritize the channel selection on the link 15$\rightarrow$20; the rest of the links choose their channels as previously, {\em given the selection on link} 15$\rightarrow$20.
For both approaches, we measure the total end-to-end throughput for the two routes.
Fig. \ref{fig:prelim} depicts these measurements %the total end-to-end throughput
for different source data rates.

\begin{figure}
\centering
\includegraphics[width=8cm]{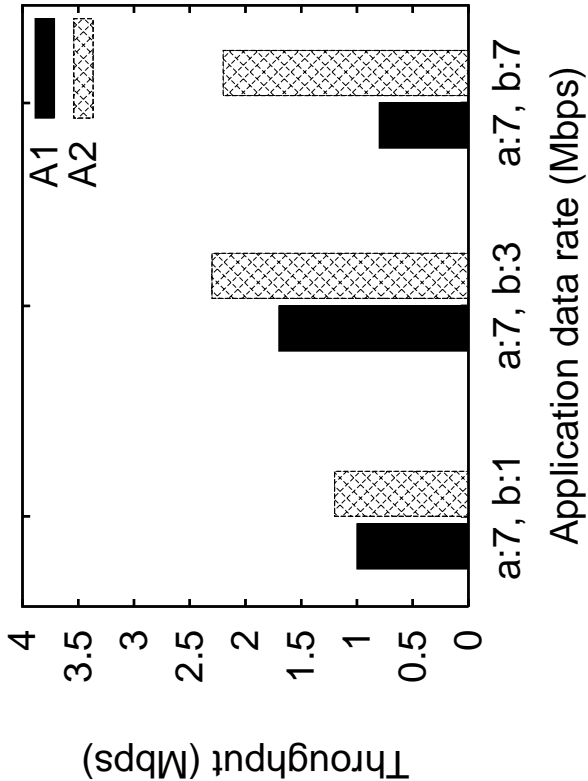}
\caption{Policy A2 outperforms policy A1, especially at high loads.} \label{fig:prelim}
\end{figure}
First, we observe that the total end-to-end throughput is always higher with $A2$. %however,
Second, we see that the throughput improvement with $A2$ is much higher than with $A1$, when the data rates of the routes are of similar magnitude.
This is because $A2$ assigns a ``better" channel to the 15$\rightarrow$20 link, which has to serve more traffic than the other links. %; it is therefore a bottleneck for the two routes.
With $A1$, however, this acts as a bottleneck to the performance of the individual routes:  this bottleneck situation cannot be captured by simply measuring the interference at each channel.
%\footnote{Most interference measurement techniques rely on measuring the RSSI of received control or data messages during a period of time \cite{powInfocom, gibbsInfocom}.}.
On the other hand,
when one of the routes, (say (a)) has significantly lower traffic demands, the aggregate  traffic that traverses link 15$\rightarrow$20 comes mainly from the route (b) over a period of time; in such a case, $A1$ works quite well, too.
This observation constitutes a key element for the design of ARACHNE, in terms of selecting channels and allocating the available interfaces to specific routes, as we explain in section \ref{sec:protocol}.

% \begin{figure}[t]
%\begin{center} \hspace{0.17in}
%\parbox{1.5in} { \hspace{0.2in}
%	\centerline{\includegraphics[width=5cm]{figures/uth.eps}}
%	\vspace{-0.18in}
%	\caption{\small The testbed deployment in the 4th (left) and the 5th (right) floor of our building.} \label{fig:ucr}
%}
%\makebox[.2in] {}
%\parbox{1.5in} {\vspace{0.1in}
%     \centerline{\includegraphics[width=2.7cm,angle=270]{figures/prelim.ps}}
%	\caption{\small Policy A2 outperforms policy A1, especially at high loads.} \label{fig:prelim}
%}
%\end{center}
%\end{figure}

The LCCS (Least Congested Channel Search) method \cite{LCCS} was the first effort towards allocating a set of available channels to wireless devices.
With LCCS, devices (e.g. APs) periodically scan the set of available channels and select the one with the lowest {\em levels of contention} (as the name suggests).
However, there are many topological scenarios where LCCS is unable to capture the total interference in the channel, as explained in \cite{sumanb}.

Similarly, Leith and Clifford \cite{clifford} propose a self-managed distributed channel selection scheme, wherein each AP passively measures the received power from the packets transmitted by neighbor APs.

Kauffmann et al. in \cite{gibbsInfocom} propose a distributed frequency selection algorithm that minimizes the global interference in the network. In this work the authors minimize the total interference which can result in improved user throughput. However, their algorithm does not consider the number of clients in the network; it assumes purely downlink saturated traffic and that all APs have affiliated clients.

Similarly, Rozner et al. in \cite{rozner}, \cite{rozner1} propose a channel assignment scheme for WLANs, taking into account traffic demands.
They show that taking into consideration the current traffic demands at APs and clients, the quality of the channel assignment can be greatly improved.
%The distributed channel hopping mechanism in \cite{maxchop},

The MaxChop mechanism \cite{maxchop} provides high levels of fairness among users using channel hopping.
%Channel hoping, h
However, it requires tight synchronization between AP and clients, while it is difficult to implement efficiently with off-the-shelf hardware.

In \cite{mobicom05} the authors study the joint channel allocation and routing problem, assuming that traffic demands and network topology are known. They %present a LP formulation of the problem and they
propose a centralized algorithm that maximizes the aggregate throughput.

Raniwala et al. \cite{hyacinth} propose a tree-based mesh architecture, Hyacinth, where the local channel  %channel
load information exchange facilitates the channel selection. %is based on this information.
Hyacinth tries to address
%In their mechanism try to solve
the joint problem of channel assignment and routing in wireless mesh networks. The latter two approaches, \cite{mobicom05, hyacinth}, however,  assume the availability of a global network view.

Our protocol differentiates from these approaches by providing efficient end-to-end channel selection in a distributed manner (access level and mesh backhaul). In ARACHNE there is no need of synchronized channel access. % in the network (impractical assumption). Besides,
Moreover our protocol does not employ %consider
any tree-based architecture, which in %many times it is not capable to
some cases cannot represent the actual network topology and its dynamics. % topology variations)
%and it adopts a new metric for load estimation. Lastly,
Finally, our work is fully compliant with 802.11s \cite{80211s} wireless mesh networks and it can be implemented on top of the existing IEEE 802.11 standards. In what follows, we present the metric that our protocol uses during channel assignment.%We compare our protocol against \emph{Hyacinth}.

%\input{metric}
%\section{A Channel Allocation Protocol for Mesh Networks}
\section{Modelling Channel Allocation}
\label{sec:theory}
It is clear that the selfish objective of the APs
is to select the channel that provides the maximum average
throughput. Therefore, the APs need to first estimate the average
throughput it can obtain from when operating on a specific channel. In this section, we investigate important parameters that need to be
taken into account for allocating the available channels to the APs in an
efficient manner, through a sophisticated metric and a problem formulation that tries to maximize the throughput in the network.

\subsection{Metric for channel selection}
 The metric that we consider in our work is called
\emph{airtime metric}.
The airtime metric was %is currently
first discussed in the 802.11s \cite{80211s} %wireless mesh networking
standard, for the purposes of load-aware routing (RM-AODV routing protocol). This metric reflects the load on a wireless router (AP) in terms of the average delay a transmission of a unit size packet experiences. RM-AODV which is the default routing protocol in 802.11s-based wireless mesh networks, employs the airtime metric in order to provide end-to-end paths with the minimum total \emph{airtime cost}.

Formally, the
\emph{airtime cost} of a link \(l\) that supports the communication of two routers, using channel or frequency \(f\), is given as:
\begin{equation}
C_{l,f} = \left[ {O_{ca}  + O_p  + \frac{{B_t }}{{R_{l,f} }}}
\right]\frac{1}{{1 - e^f_{pt} }}.\label{eq_airtime}
\end{equation}
In \eqref{eq_airtime},
\( O_{ca}\) is the channel access overhead,
\(O_p\) is the protocol overhead and
\(B_t\) is the number of bits in the test frame\footnote{The transmission of test frames is necessary, in order to derive values for the computation of the airtime cost.}.
Some representative values for these constants, for 802.11g, are:
\(O_{ca} + O_p=1.25\)ms and
\(B_t=8224\)bits. Furthermore, %The input parameters
$R_{l,f}$ and \(e^f_{pt}\) are the current transmission rate and frame-error rate, respectively,  in
%bit rate in \(Mbs\),
Mbps,
%and the frame error rate
for the test
frame size \(B_t\) in channel $f$. %, respectively.
%The rate $r^i$ depends on the
%local implementation of rate adaptation, and represents the rate at
%which the mesh point would transmit a frame of standard size \(B_t\)
%based on the current conditions.
In other words, %The
the
estimation of \(e^f_{pt}\) %is
%also a local implementation choice; %and
%it is intended to estimate
%\(e_{pt}\) is estimated
corresponds to
%for
transmissions of standard-size frames \(B_t\) at the
current transmit bit rate \(R_{l,f}\).

In order to understand the reasoning behind the airtime metric, we
should refer to the recent work investigating the calculation of the
average throughput of the 802.11 based WLANs \cite{bianchi},
\cite{gupta}, \cite{hidden}, \cite{kumar}. It has been observed that
when there are several flows with different physical transmission
rates, then the throughput of all flows is bounded by the slowest
transmission rate \cite{anomaly}.  In order to explain this anomaly,
further studies have been conducted, and the average uplink
throughput in a single cell environment is calculated under
saturation and decoupling approximations in \cite{bianchi} and
\cite{kumar}. The saturation approximation states that there are
always packets backlogged on every user.  Meanwhile, with the
decoupling approximation it is assumed that when there are $n$
users, the aggregate attempt process of $(n-1)$ nodes is independent
of the back-off process of any given node.

We consider the simple case, when all nodes have the same back-off
parameters, each node is the transmitter for a single flow, and all
packets have lengths are equal to $L$.  As derived in \cite{kumar},
the total average network throughput $\theta(\beta)$ is given as:

\footnotesize
\begin{flalign}
&\theta(\beta)=&\nonumber\\ 
&\frac{n\beta(1-\beta)^{n-1}L}{1+\sum_{i=1}^n\beta(1-\beta)^{n-1}(\frac{L}{C_i}+T_0)+(1-(1-\beta)^{n}-n\beta(1-\beta)^{n-1})T_c}&
\end{flalign}
\normalsize
where $\beta$ is the attempt rate (probability) in the equilibrium,
$C_i$ is the physical transmission rate of node $i$, $T_0$ is the
fixed overhead with packet transmission and $T_c$ is the fixed
overhead for an RTS collision.  Due to the exponential back-off
behavior of the nodes and the decoupling approximation, it can be
shown that the attempt probability of a node accessing the channel
can be determined in terms of a given collision probability $\gamma$
as:
\begin{align}
G(\gamma)=\frac{\sum_{k=0}^K \gamma^k}{\sum_{k=0}^K \gamma^k
b_k},\label{eq:beta}
\end{align}
where $K$ is the maximum number of attempts allowed under the
protocol, and $b_k$ is the mean back-off at the k$^{th}$ attempt.
Meanwhile, the probability of collision of an attempt by a node is
given by $\Gamma(\beta)=1-(1-\beta)^{n-1}$ due to the decoupling
approximation. The equilibrium behavior of the system is governed by
the solution of the fixed point equation $\gamma=\Gamma(G(\gamma))$.
If this equation is solved it yields the collision probability from
which the attempt rate in the equilibrium $\beta$ can be determined
from \eqref{eq:beta}.

Note that $\frac{L}{\theta(\beta)}$, is the average delay per packet
in the equilibrium, which is given as:
\begin{align}
\rho(n)=\frac{L}{\theta(\beta)}=\frac{1}{n\beta(1-\beta)^{n-1}}+(T_0-T_c)\nonumber\\ +\frac{1-(1-\beta)^{n}}{n\beta(1-\beta)^{n-1}}T_c+\frac{1}{n}\sum_{i=1}^n\frac{L}{C_i}
\label{eq:airtime}
\end{align}
The first three terms in \eqref{eq:airtime} represent the delay due
to channel contention and protocol overheads, and the last term
represents the average transmission time of an $L$ length packet by
a node in the cell.

\begin{figure}
\centerline{\includegraphics[angle=0,width=3in]{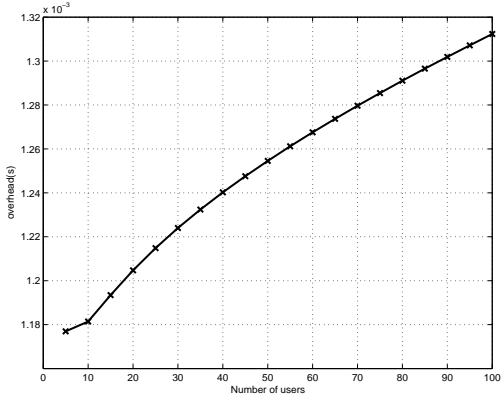}}%\vspace{-0.2in}
\caption{The contention and protocol overhead calculated according
to \eqref{eq:airtime}.} \label{fig:overhead}
%\vspace{-0.3in}
\end{figure}

Comparing the definition of the airtime metric $C_{l,f}$ and
$\rho(n)$, we can see that the sum of $O_{ca}$ and $O_p$ correspond
to the approximation of the first three terms in \eqref{eq:airtime}
with a constant value.  Clearly, the transmission of RTS, CTS, DATA
and ACK frames may also get corrupted not only due to collisions but
also due to channel errors.  The authors in \cite{gupta} extended
the analysis in \cite{bianchi} to account for wireless bit errors.
The goodput expression in \cite{gupta} assumes that the wireless
channel can be modeled by an appropriate Gilbert model with known
transition probabilities.  In wireless networks with dynamically
changing conditions, such an assumption is not practical. Therefore,
in our method, the current frame error rate, $e^f_{pt}$ is measured by
the users and AP.  For each packet attempted to be transmitted, it
would be in error due to channel errors with probability $e^f_{pt}$.
Clearly, the average number of attempts until successful
transmission would be $1/(1-e^f_{pt})$.  For each attempt on the
average $\rho(n)$ amount of time is experienced.  The product of
$1/(1-e^f_{pt})$ and $\rho(n)$ gives the airtime metric used in our
algorithm.

When the overhead is calculated as described in \eqref{eq:airtime},
we observe that it varies with the number of nodes in the cell.  The
overhead with respect to the number of nodes is given in Figure
\ref{fig:overhead}, when $K=7$, $b_0=16$, $b_k=2^kb_0$, the slot
length is $20\mu s$, $T_0=52$ and $T_c=17$ slots.
The overhead varies between 59 and 66 slots, and
thus, it is clear that there is weak dependency between the number
of users $n$ and the observed overhead. Therefore, in our work
the overhead is taken as a constant to avoid additional
computational complexity.

\subsection{Optimal channel allocation problem formulation}
In this section we give a general formulation of the channel allocation in 802.11-based multi-radio wireless networks. We formulate the problem of optimal allocating the available channels, assuring that the maximum path end-to-end delay (total airtime cost of the links that are part of the path) in the network is minimized. Based on our previous analysis this will provide throughput maximization and balanced/stable network operation is assured \cite{bertsekas}:

%\[\begin{array}{c}
%\mathop {\min }\limits_{{a_{l,f}}} \mathop {\max }\limits_{p \in P} \sum\limits_{l \in L_p} {\sum\limits_{f \in F} {a_{l,f}{C_{l,f}}}}, \\
%subject\begin{array}{*{20}{c}}
%{}
%\end{array}to:\begin{array}{*{20}{c}}
%{}&{}&{}&{}&{}&{}&{}&{}&{}&{}&{}&{}&{}&{}&{}&{}&{}&{}
%\end{array}\\
%{a_{l,f}} = [0,1],\forall l \in L_p, \forall p \in P, \forall f \in F\\
%\sum\limits_{f \in F} {{a_{l,f}} = 1,} \forall l \in L_p, \forall p \in P\\
%{C_{l,f}} = \left[ {O_{ca}  + O_p  + \frac{{B_t}}{{R_{l,f} }}}
%\right]\frac{1}{{1 - e^f_{pt} }}, \forall l \in L_p, \forall p \in P, \forall f \in F
%\end{array}\]

\begin{subequations} \label{objfunc_flow}
\begin{flalign}
\mathop {\min }\limits_{{a_{l,f}}} \mathop {\max }\limits_{p \in P} & \sum\limits_{l \in L_p} {\sum\limits_{f \in F} {a_{l,f}{C_{l,f}}}}  \\
{\rm s.t.} \quad & {a_{l,f}} = [0,1], \quad \forall l \in L_p, \quad \forall p \in P, \quad \forall f \in F,\\
& \sum\limits_{f \in F} {{a_{l,f}} = 1,} \quad \forall l \in L_p, \quad \forall p \in P,\\
& {C_{l,f}} = \left[ {O_{ca}  + O_p  + \frac{{B_t}}{{R_{l,f} }}}
\right]\frac{1}{{1 - e^f_{pt} }}, \quad \forall l \in L_p,\nonumber\\ 
& \forall p \in P, \quad \forall f \in F
\end{flalign}
\end{subequations}

Where $P$ is the set of active paths in the network, $F$ is the set of available channels/frequencies that could be used by the APs, $L_p$ is the links that belong to path $p$, $a_{l,f}$ is the probability that frequency $f$ is used in the communication through link $l$, and $C_{l,f}$ is the airtime cost of link $l$ that operates in frequency $f$ .

%\section{A Channel Allocation Protocol for Mesh Networks}
\section{Dynamic Channel Selection Protocol}
\label{sec:protocol}

Inspired by the previous problem formulation and the \emph{airtime metric} association with the average throughput in the network, we present ARACHNE, a load and routing aware channel selection protocol.
The design of ARACHNE makes the following assumptions: a) Nodes of the mesh backhaul are equipped with at least 3 radios (2 for the mesh backhaul communication and 1 for the access level), one of which is set to a pre-arranged channel, $c_p$, which  is the same for all nodes in the network, b) A client associates to the AP with the strongest signal (RSSI) among all neighbor APs (802.11-based association procedure), c) The set of channels used at the access level (AP-client interfaces), is different from the channel-set used at the mesh backhaul. For simplicity in this paper, the AP-client interfaces operate in the 2.4 Ghz band; the 5 GHz band is used exclusively for the interfaces in the mesh backhaul, and d) We assume that traffic is exchanged among end hosts belonging to the same mesh network, i.e., traffic is not crossing different networks.

%Our analysis from the previous section clearly shows that the airtime metric reflects the performance of the network in terms of throughput.
%In other words, determining the channel with the lowest airtime cost can provide the locally maximum average long-term throughput.
%This throughput for a channel $a$ is given by:
%
%\begin{equation}
%\theta_a (\beta ) = \frac{1}{{(1 - e_{pt} )}}\frac{L}{{C_a }}
%\label{eq:airtime3}
%\end{equation}
%
As previously mentioned, ARACHNE is using
\emph{airtime metric} that is an approximation of the per packet latency. %in the current communication channel.
The airtime metric was %is currently
first discussed in the 802.11s \cite{80211s} %wireless mesh networking
standard, for the purposes of load-aware routing (RM-AODV routing protocol). This metric reflects the load on a wireless router (AP) in terms of the average delay a transmission of a unit size packet experiences. RM-AODV which is the default routing protocol in 802.11s-based wireless mesh networks, employs the airtime metric in order to provide end-to-end paths with the minimum total \emph{airtime cost}.
We adopt this metric in ARACHNE for the purposes of our proposed channel selection functionality.

Several
 studies have shown that %the interference affects the performance of the communication in 802.11 wireless networks
the number of erroneously received packets %packet-dropping
increases and the transmission rate decreases when the cells in the network interfere
 \cite{niculescu}, \cite{niculescu1}, \cite{gibbsInfocom}.
%Therefore, we must capture these effects in order to detect and face the interference problem in our protocol.
%The
The \emph{airtime metric} takes into account the packet error rate  %packet-dropping
as well as the transmission rate; % and so,
hence,
it reflects the performance %of
at a particular
 communication channel. % in the network.
 Besides, our analysis shows that the average \emph{airtime cost} is an approximation of the average per-packet delay. Therefore, the average \emph{airtime cost} is a representative metric that reflects the channel performance and also approximates the maximum throughput in the network.

ARACHNE captures the performance of
a bidirectional link in terms of estimated throughput at a particular channel,  %, when the AP operates in
%the available channels that can be used by the APs in the network,
by
measuring the average \emph{airtime cost} for both uplink and downlink,  $C_{l,f} = \overline {C^{up}_{l,f}}+\overline {C^{down}_{l,f}}$ (\emph{airtime cost} for a particular link \(l\) that operates on channel \(f\)), and applies a channel selection methodology where the
channel with the
minimum $C_{l,f}$ is chosen.
This channel selection policy determines the frequency with the minimum  average per-packet delay in both uplink and downlink, % channels
%and therefore, it
%provides the
%approaches the
thereby approximating the
maximum throughput in the communication. The goal of ARACHNE is to assign channels to mesh nodes (their associated links), such that the %locally-computed
average airtime metric is minimized, both at the access and the backhaul levels.
Hence, ARACHNE involves two procedures, P1 and P2, one for each level.
ARACHNE is executed exclusively by the nodes that belong to the mesh backhaul, i.e., the relay nodes as well as the APs that connect the end-hosts with the mesh network.
The channel discovery is initiated by the AP of the source host.
In what follows, we describe the operations of the protocol, P1 and P2, for each of the two levels of operation.

\subsection{Channel Selection at the Access Level \textbf{(P1)}}
The access level involves the communication of the end hosts (clients) with mesh nodes at the edge of the backhaul (APs).
Let us assume that host $A$ wishes to send traffic to another host $B$. This traffic will flow through $A$'s AP, $M_A$, to the mesh backhaul and it will eventually reach $M_B$; the latter will finally forward the traffic to $B$.
With ARACHNE, the frequency selection at the access level involves a channel discovery process at the two aforementioned APs, in order to find the channel with the minimum airtime cost value (we provide the steps of this process below).
The calculation of this value is performed for every scanned channel, and involves all the downward and upward links of $M_A$, $M_B$.
Let us again assume that host $A$ wants to send traffic to host $B$.
%The channel selection process at the access level is represented by the flow diagram of Fig. \ref{fig:accesslevel}
%\begin{figure}[t]
%\centering
%\includegraphics[width=2in]{figures/Flow2.eps}
%\caption{Channel discovery/selection at the access level.}
%\label{fig:accesslevel}
%\end{figure}
%%
This part of the protocol, P1, is executed by the APs at the access level, and consists of the following steps.

\begin{enumerate}

\item{\noindent\textbf{[P1a]. Deriving interference information for the current channel.}
At the nominal start of ARACHNE,  $M_A$ is informed about the operational frequencies of its neighbor {\em co-channel}  APs.
%(This information is crucial for the calculation of the airtime cost, as per Eq. \eqref{eq:cumairtime}).
This can be performed through passive scanning of periodically transmitted beacons \cite{mdg}, \cite{mdg1}.
%In this work, we follow the former approach.
By the end of this step, $M_A$ is aware of the total received power from all co-channel APs.}

\item{\noindent\textbf{[P1b]. Computing the downlink airtime cost.}
$M_A$ calculates the aggregated downlink airtime cost with its clients, through a link performance-measurement procedure, described in detail in \cite{Athanasiou} where airtime metric is used for user association.
%$M_A$ further embeds the computed cost into its
%link measurement-request transmissions \cite{80211k}
%beacon frames, in order for its clients to compute the uplink costs.}

\item{\noindent\textbf{[P1c]. Computing the uplink airtime cost.}
The clients of $M_A$ %read the downlink airtime cost from $M_A$'s message, and further
calculate their individual uplink costs.
%They include these costs into their measurement report messages towards $M_A$.
They piggy-back this information
through their data frame transmissions towards
$M_A$. By the end of this step, $M_A$ has received
information regarding the uplink channel qualities from all
its clients.}

\item{\noindent\textbf{[P1d]. Deciding if the current channel is appropriate.}
$M_A$ receives the information from the client and computes the average airtime cost (for both uplink and downlink) for the access level.
If this is higher than a pre-defined threshold $T$, $M_A$ remains in the same channel; otherwise it initiates a channel discovery process. The value of threshold $T$ is decided and controlled by the system designer.}

\item{\noindent\textbf{[P1e]. Computing the cumulative airtime cost at the next available channel.}
%$M_A$ forces its clients to switch to the next channel, through a certain control frame\footnote{The measurements on our testbed indicate that this channel switching of the AP and its clients is performed quite quickly.}.
%%Yannis: which type of frame?
%$M_A$ repeats the steps $P1a-P1e$ for the new channel.
%\textbf{\em If all available channels have been visited, $M_A$ finally selects the %channel with the minimum cumulative airtime cost. }
$M_A$ and its clients switch to
the next channel and repeat steps $P1a-P1e$.%\footnote{The measurements on our testbed indicate that this channel switching of the AP and its clients is performed quite quickly.} If all available
channels have been visited, $M_A$ finally selects the channel
with the minimum average airtime cost (for both uplink
and downlink).}}
\end{enumerate}

%More information about the calculation procedure, the threshold $T$ value (which determines the frequency of invocation of the above procedure) and the convergence duration of ARACHNE can be obtained in \cite{Tech_report_channel}.
Note here that process P1 (of selecting a channel at the access level) is actually independent of P2; although they are executed in parallel, they do not  affect each other to a large extent, in terms of convergence time, since they utilize different sets of channels.

\subsection{Channel Selection at the Mesh Backhaul \textbf{(P2)}}
The role of
the mesh backhaul is to serve the forwarding of packets towards their final destinations.
%In other words, nodes of the mesh backhaul act as relays of the packets exchanged between the end hosts.
%An important part of our end-to-end protocol is the channel allocation at the mesh backhaul communication.
%Usually MAPs are equipped with 2-3 wireless interfaces and then the problem of assigning the available channels to these interfaces arises.
Undoubtedly, the network connectivity is affected by a potential channel selection policy that  may be applied in the mesh backhaul. %Therefore, channel allocation must take
ARACHNE's channel selection framework ensures
connectivity at all times in the network, while the selection of a frequency at a particular link takes into account the load of the routing sessions that are traversing the link.
%In this way several unexpected situations will be avoided (routing session breaking, disconnected graphs, etc.)

%\noindent{\bf Load-aware routing with RM-AODV:}
For the purposes of this study, we assume that the RM-AODV routing protocol \cite{80211s} is applied on the mesh backhaul. RM-AODV is the default routing protocol proposed in the context of 802.11s mesh networks, where the airtime cost is used as a routing decision metric in the mesh backhaul. In particular, the airtime metric is appended to the RREQ and RREP messages, during the route discovery process; finally, the end-to-end path with the minimum total airtime cost (the route that has the minimum load) is selected. Our choice of RM-AODV is motivated by the fact that this routing protocol
is based on the airtime metric.
{\em Note, however, that ARACHNE can operate in conjunction with {\bf \em any} load-aware routing protocol!}

%\noindent{\bf Control information exchange using LABA:}
ARACHNE employs the Local Association Base Advertisement (LABA) mechanism introduced in 802.11s \cite{80211s}, in order to
disseminate information with regards to
inform  the entire mesh network about the
clients (end-hosts) that are associated with each mesh AP (MAP).
MAPs periodically broadcast LABA frames, which consist of the MAC addresses of the hosts that are associated with. %Therefore, the MAPs in the network obtain the information that they need in order to perform different functionalities (like routing discovery to a destination, etc.). It is obvious that this mechanism plays an important role in the operation of a mesh network
%the MAPs.
We have enhanced the LABA frames to include load related information, as we explain later in the description of the main parts of our protocol.
Our protocol prioritizes the channel assignment on the most loaded mesh APs, as well as on the mesh APs that
are expected  %/estimated
to be highly loaded in the near future.
In particular, ARACHNE observes the per-link load, for the links that serve one or more routes in the mesh;
the load is captured in the airtime cost metric.
%The RM-AODV protocol informs ARACHNE about the routing sessions that are active and their load. As we have previously mentioned the CCM coordinates the channel negotiation. In our protocol we suppose that the mesh APs are equipped with more than 2 interfaces.
The main problem in designing distributed channel selection policies is the channel dependencies that arise between the nodes that are part of the mesh network. For example, we assume that in the simple network in Fig. \ref{example} nodes are equipped with 2 wireless interfaces. Node D uses channel \emph{a} in order to communicate with node E and node E uses channel \emph{b} to communicate with nodes F and G. In case that in the link E-F the current channel is heavily loaded, node E uses a new channel \emph{c} that operates better at the link E-F. As E has only two interfaces, channel \emph{c} must be used at the link E-G too. Unfortunately, a strong dependency between links E-F and E-G is established. Channel \emph{c} may not operate effectively at the link E-G and therefore the performance of the network is decreased. The aforementioned \emph{ripple effect} could be further propagated in dense/huge mesh networks. In order to avoid {\em ripple effects} in the channel selection process \cite{hyacinth}, we categorize the wireless interfaces at each mesh AP that serves as a relay as:
\begin{itemize}
\item \emph{ {\bf \em IN} network interfaces} that are used for data reception, % and
\item \emph{ {\bf \em OUT} network interfaces} that are used for data forwarding.
\end{itemize}
%{\bf **** SAY MORE DETAILS HERE ****}
In addition, each node can assign channels only to its \emph{\textbf{OUT}} interfaces. The \emph{\textbf{IN}} interfaces follow the channels that are assigned by the nodes that communicate with the current node. In our protocol we assume that each node is equipped with at least one \textbf{\emph{IN}} and one \emph{\textbf{OUT}} interface.
%\textbf{**** I DONT LIKE THIS, I THINK WE MUST OMIT THIS ****}
%(Recall here that, each mesh node has set a specific, pre-arranged channel $c_p$, on one of its available interfaces.
%With this, ARACHNE guarantees that the network remains fully connected at all times.)

%\noindent\textbf{Routing-aware channel allocation:}
In what follows,
we describe the  %basic
steps that are executed  %in our routing-aware channel allocation protocol:
by ARACHNE, for assigning channels at the mesh backhaul (P2).
To begin with, we consider a network state, in which:
(a) All nodes have already dedicated one of their\emph{\textbf{ OUT}} interfaces ($I_p$), %$I_p$,
to control channel $c_p$;
(b) The RM-AODV protocol has converged to a set of  routes between end-hosts;
(c) All of the other wireless interfaces of a mesh AP (besides $I_p$) have been randomly assigned a channel in the 5 GHz band.
The channel assignment in P2 is comprised of the following steps.
%\begin{itemize}
%\item
%{\textbf{STEP 1}: Initially a common channel must be applied in two of the wireless interfaces at each mesh AP (IN and OUT), in order to achieve full connectivity in the network. In case that there are more available interfaces, the mesh AP randomly assigns channels to them. Consequently, end-to-end connectivity is guaranteed and special situations are avoided (like disconnected graphs, etc.). RM-AODV is applied on top of this initial channel allocation and finds the optimal paths according to the end-to-end airtime cost calculation.}
%\item

\begin{enumerate}
\item {\noindent{\textbf{[P2a]. Constructing a priority list.}
With ARACHNE, an AP starts the channel-scanning process at a time-instant dictated by its priority ranking.
This priority is highly-related to the load of each mesh AP; the higher the load (the data that must forward), the higher the priority. In addition, we believe that in this prioritization the estimated load of a mesh AP in near future must be taken into account. In this way we guarantee that our protocol will converge to a long-lasting and a stable channel allocation.
Hence, it is imperative that this list is constructed, before APs start scanning each channel.
Each mesh AP $a$ calculates its priority rank according to its current load and its expected load, as:
$$P_r^a=L_{crnt}^aw_1+L_{est}^aw_2,$$
where
\(L_{crnt}^a\) is the current load,
\(L_{est}^a\) is the estimated load
and
\(w_1\), \(w_2\) are the weights that are used in the calculation (we will give more details about the selection of \(w_1\), \(w_2\) in the evaluation of our protocol).}}

\noindent As far as the calculation of the estimated load \(L_{est}^a\) is concerned, we adopt the estimation method (based on historical data), proposed in \cite{infocom08}. In this approach the authors design a trace-based traffic model in order to predict the aggregated traffic demands of an AP in near future. Time-series analysis is the basis of this traffic estimation model. The accuracy of this model is high and in combination with our load-aware channel selection protocol we achieve long-lasting and stable channel allocation in the mesh network.

\noindent Besides, the priority ranks are incorporated into the LABA frames and are broadcasted in the network.
Consequently, a sorted list (in terms of channel selection priority rank) is  %achieved
disseminated and maintained by all APs.
Hence, by the end of this step each AP knows the priority of all APs in the network.

%\item
\item{\noindent{\textbf{[P2b]. Performing channel-scanning.}
Let us assume that each AP can scan a set of $K$ channels.
The first mesh AP in the priority list  %executes the following procedure (supposing that we have K available channels): 1) In case that it is equipped with one OUT wireless interface, it must find the channel that will optimize the performance of the flows F that are has active in its' outgoing links. Therefore, it
measures the cumulative airtime cost for each
of the $K$ channels, and constructs a local, {\em temporary},
sorted list with the cumulative airtime cost, per channel.}}

\item{\noindent{\textbf{[P2c]. Assigning route sessions to available \textbf{\em OUT} interfaces.}
Each mesh AP typically prefers to assign low-airtime channels to links that serve high-load end-to-end traffic sessions. %, taking into account the saturated throughput, per channel, under the current channel conditions (captured by the airtime cost value).
%As we explained in section \ref{sec:prelimexp} (see example with the 21$\rightarrow$35 link), the bandwidth-sharing at a particular channel provides higher end-to-end throughput when the served sessions have different application data rate requirements.
ARACHNE manages to efficiently utilize the available spectrum, by providing a balanced channel and interface assignment in the network.
%allocating channels/interfaces to diverse, in terms of traffic volume, sessions .
A mesh AP $a$ calculates the maximum load \(L^a_{sh}\) that can be assigned at each of its $OUT$ interfaces, as (clearly, when the number of available interfaces is higher than the number of served sessions, ARACHNE allocates an interface to a particular session):
$L^a_{sh}=L^a_{crnt} / {N^a_{OUT}},$
where \(N^a_{OUT}\) is the number of the available $OUT$ interfaces of mesh AP $a$, and $L^a_{crnt}$ is the current load that must be served by the $OUT$ interfaces.
%Moreover, the AP computes the cumulative airtime cost of the flows that are active in each link, and
%further performs a load-balancing interface-assignment strategy where the main constraints are:
The main constrains of the interface-assignment strategy are:
(a) The load assigned to an $OUT$ interface is less than \(L^a_{sh}\),
and
%(b) the cumulative airtime cost of the flows assigned to each %link
%interface is optimized according to the channel that we have selected using our channel allocation policy.
b) The load is proportionally allocated to the available $OUT$ interfaces in terms of the load of the flows that pass through the current mesh AP. In other words, the load is balanced to the available $OUT$ interfaces.}}

\item{\noindent{\textbf{[P2d]. Assigning channels to \textbf{\em OUT} interfaces.}
In order to assign a channel to an $OUT$ interface, a coordination with the $IN$ interface of the mesh AP that receives the traffic is required.
%We apply CCM \cite{citationforCCM} to achieve this coordination,  %between the affiliated interfaces,
%as we explained earlier.
%\noindent{\bf **** YANNH CHECK THIS, WE WILL DISCUSS IT ****}
In particular, a mesh AP \emph{A} that selected the channel with the minimum airtime cost in the communication with a mesh AP \emph{B}, sends an RTC frame to \emph{B} announcing in this way that a new channel must be used in their communication. Then, \emph{B} responds with a CTC frame and sets one of its \emph{IN} interfaces to the selected channel. In case that \emph{B} hasn't responded in a time window (timeoff), \emph{A} retransmits its RTC frame. We must note here that there are special situations where \emph{B} has recently assigned channels to its \emph{IN} interfaces (during the same protocol iteration) and there are no available interfaces to assign the channel proposed by \emph{A}. In other words, higher priority APs (compared to the priority of \emph{A}) have previously assigned those channels to \emph{Bs'} \emph{IN} interfaces and there are no available \emph{IN} interfaces to use. In this case \emph{B} responds with an XTC frame announcing this situation to \emph{A} and the available channels that are assigned to its \emph{IN} interfaces. Lastly, \emph{A} is restricted to use one of those channels in the communication with B (the channel with the minimum airtime cost in the link A-B).
%Note that this process is performed for each of the $K$ channels.
After the completion of the aforementioned process, the AP sends a  ``good-to-go"  unicast\footnote{This unicast message speeds up the process of completing a full iteration of channel scanning for all participating  mesh  APs.} message to the next AP in the list, to initiate its channel scanning and assignment process.}}

%\item
\item{\noindent{\textbf{[P2e]. Selecting channels iteratively.}
%The next AP in the sorted list performs the aforementioned procedure, after receiving the  ``good-to-go"  unicast message from the previous AP.
When all APs have completed a round of channel scanning,
RM-AODV updates the routing tables at the mesh backhaul, taking into account the new
state of the network.
Note that RM-AODV discovers routes based on the computation of the airtime cost, which will have likely changed after an iteration of the channel selection process.
Steps P2a-P2e are repeated until the channel selection has converged.}}

\end{enumerate}

\section{Evaluating ARACHNE}
\label{sec:evaluation}

In this section we present the evaluation study of our approach. During the first part of our validation methodology we use the optimal solutions (resulted from solution of the channel allocation problem described in section \ref{sec:theory}) as a benchmark in the direction of evaluating ARACHNE. We are using IBM ILOG CPLEX Optimizer \cite{ilog} to find the optimal solutions.
%our load-aware channel allocation protocol. We adopt an evaluation methodology where both
Moreover, we evaluate ARACHNE through extensive % OPNET
simulations, which import both
 synthetic \cite{opnet} %traces
and real traces \cite{dartmouth}, \cite{ibm}. % are used.
%Firstly, we simulate our protocol in OPNET environment in order to give a perception of the behavior of the proposed mechanisms under different operational conditions and application demands. In addition, we use real traces to realistically evaluate the performance of our protocol and compare it with other channel allocation protocols in literature.
We compare our protocol against other channel selection schemes,
and we present ARACHNE's predominance in terms of the total network throughput, average packet dropping and average transmission delay.

%\subsection{OPNET-based traffic demands}
%\noindent\textbf{Simulation set-up:}
We have implemented ARACHNE
in OPNET \cite{opnet}. %, taking into account the 802.11 protocol operations.
%, using the basic procedures defined by the IEEE 802.11 standard.
%We have modified the {\em beacon} frames of 802.11
%{\bf **** SAY EXACTLY WHICH FRAMES WE HAVE MODIFIED ****} %,in order to incorporate
%to facilitate the information exchange and distribution that our protocol design requires.
%
%elements that our system requires. %needs.
%Note that we have not changed the packet synchronization procedure,  %and the small modifications we added for our mechanisms,
%and that we have verified that our modifications do not have a negative impact on the overall performance of the 802.11 protocol in terms of overhead. %network.
%
%The nodes that consitute the mesh backhaul are the %represented by the
%We have simulated a wireless mesh network in the OPNET simulation environment.
%The
%wireless routers that are provided by the OPNET wireless module. % are part of the backhaul network.
%The peripheral routers of the mesh, serve as the APs that have direct links with the end hosts (clients). %as well.
We have also
implemented the RM-AODV protocol and we consider this routing protocol in our simulations. % that is introduced by 802.11s
%\cite{80211s} standard and we have applied this routing protocol at the mesh backhaul. The STAs
 The clients are uniformly distributed (at random) in the 1000m$\times$1000m simulation area. All nodes use a default transmit power of 20 dBm and the source-destination pairs are randomly chosen in the network. We experiment with: (a) fully-saturated, end-to-end UDP traffic, (b) VoIP traffic and (c) real traffic traces from Dartmouth College \cite{dartmouth} and IBM \cite{ibm}.
 The weight values that are used in our simulation experiments are: $w_1=0.6$ and $w_2=0.4$ (the current load affects the channel selection procedure more than the estimated load in near future). As far as the duration of the measurement period is concerned, in our experiments the APs remain $100ms$ in each scanned channel in order to gather the necessary information. The convergence of our mechanisms is reached rather quickly in our network topologies, as will be presented later in the current section (close to $20sec$ in average and after a small number of iterations). Moreover, the traffic keeps flowing during the execution of ARACHNE and the network operation is not affected. We chose accordingly the values of the performance threshold $T$ in order to avoid unnecessary and frequent protocol executions.
%{\bf ***** I REMOVED TEXT HERE -- CHECK THE .TEX TO SEE IF YOU AGREE ******}
%{\bf ***** I DON'T UNDERSTAND WHY YOU FIX THE DATA RATE AND THE BIT RATE, AND NOT USE SATURATED-AUTORATE. WE WILL RECEIVE MANY COMPLAINS ABOUT THIS. I SUGGEST WE REMOVE IT. *****}
%For the communication between the wireless routers in the backhaul network,
%we use the %physical
%model of IEEE 802.11a OFDM %physical
%PHY layer.
%The supported physical rate is 12 Mbps (12 communication channels are available in the mesh backhaul). The STAs that are associated with the available peripheral APs transmit their packets at 1 Mbps. We vary the number of source/destination pairs in order to vary the overall load. The source and destination nodes are chosen randomly among the nodes in the network.
Throughout our simulation experiments we
compare the network performance with ARACHNE, against the single-channel assignment, a random-channel allocation strategy, as well as
the Hyacinth protocol \cite{hyacinth}.

\begin{table}[t]
\caption{Performance with UDP traffic and $12$ orthogonal channels available} {
\begin{center}
\footnotesize
\begin{tabular}{|c|c|c|c|}
  \hline
  \textbf{Number} & \textbf{Optimal } & \textbf{Throughput } & \textbf{Average } \\
  \textbf{of APs} & \textbf{Throughput} & \textbf{achieved by} & \textbf{ARACHNE}\\
  \textbf{} & \textbf{(Mbps)} & \textbf{ARACHNE (Mbps)} & \textbf{iterations}\\
  \hline\hline
  $10$ & $48.2$ & $45.1$ & $3$  \\
  \hline
  $20$ & $79.4$ & $76.4$ & $5$  \\
  \hline
  $30$ & $131.2$ & $121.3$ & $8$  \\
  \hline
  $40$ & $164.8$ & $151.1$ & $11$  \\
  \hline
  $50$ & $208.6$ & $192.7$ & $15$  \\
  \hline
  $60$ & $249.1$ & $230.2$ & $19$  \\
\hline
\end{tabular}
\end {center}}\label{tableUDP11}
\end{table}
\begin{table}
\caption{Performance with UDP traffic and $3$ orthogonal channels available} {
\begin{center}
\footnotesize
\begin{tabular}{|c|c|c|c|}
  \hline
 \textbf{Number} & \textbf{Optimal } & \textbf{Throughput } & \textbf{Average } \\
  \textbf{of APs} & \textbf{Throughput} & \textbf{achieved by} & \textbf{ARACHNE}\\
  \textbf{} & \textbf{(Mbps)} & \textbf{ARACHNE (Mbps)} & \textbf{iterations}\\
  \hline\hline
  $10$ & $42.7$ & $38.4$ & $2$  \\
  \hline
  $20$ & $64.8$ & $59.2$ & $3$  \\
  \hline
  $30$ & $112.2$ & $102.2$ & $6$  \\
  \hline
  $40$ & $139.1$ & $121.7$ & $9$  \\
  \hline
  $50$ & $194.2$ & $172.4$ & $11$  \\
  \hline
  $60$ & $227.5$ & $211.8$ & $15$  \\
\hline
\end{tabular}
\end {center}}\label{tableUDP3}
\end{table}

\begin{figure}[t]
\centering
\subfigure[Total throughput Vs. \# clients.]{
\includegraphics[width=2.5in]{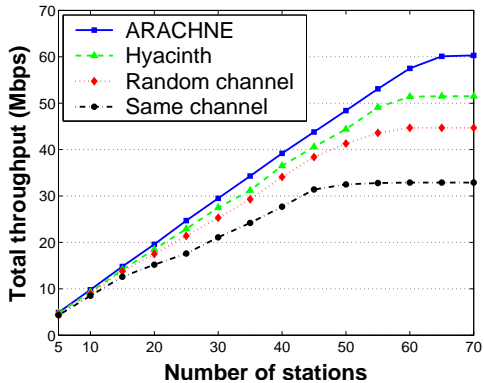}\label{cbr_sta}}\\
\subfigure[Total throughput Vs. \# APs.]{
\includegraphics[width=2.5in]{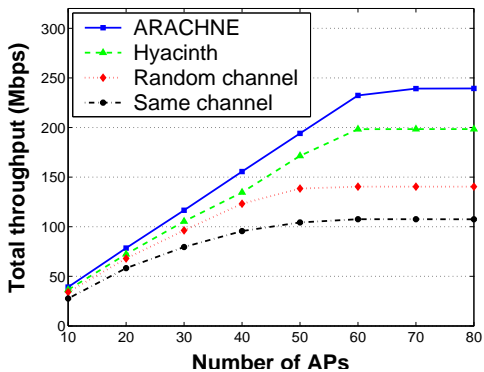}\label{cbr_ap}}
\caption{ARACHNE povides very high total end-to-end throughput with saturated UDP traffic.}
\label{fig:cbr} %\vspace{-0.3in}
\end{figure}
%We measure the total achieved network throughput, the average end-to-end delay and the average dropped data packets in the network.
%
%\begin{figure}[t]
%\centering
%\subfigure[Total throughput Vs. \# clients.]{
%\includegraphics[width=1.5in]{figures/throughput_sta.eps}\label{cbr_sta}}
%\subfigure[Total throughput Vs. \# APs.]{
%\includegraphics[width=1.5in]{figures/throughput_ap.eps}\label{cbr_ap}}
%\caption{\small ARACHNE povides very high total end-to-end throughput with saturated UDP traffic.}
%\label{fig:cbr} %\vspace{-0.3in}
%\end{figure}
%
%\begin{figure}[t]
%\centering
%\subfigure[Average end-to-end delay.]{
%\includegraphics[width=1.5in]{figures/end_to_end_delay.eps}\label{voip_delay}}
%\subfigure[Average dropped data.]{
%\includegraphics[width=1.5in]{figures/dropped_data.eps}\label{voip_dropped}}
%\caption{\small VoIP simulation results.}
%\label{fig:voip} %\vspace{-0.3in}
%\end{figure}

%
%
% \begin{figure}[t]
%\begin{center}
%\parbox{1.5in} {
%	\centerline{\includegraphics[width=4.5cm]{figures/ucr.eps}}
%	\caption{\small Testbed deployment.} \label{fig:ucr}
%}
%\makebox[.12in] {}
%\parbox{1.5in} {
%     \centerline{\includegraphics[width=3cm,angle=270]{figures/prelim.ps}}
%	\caption{\small Policy A2 outperforms policy A1, especially at high loads.} \label{fig:prelim}
%}
%\end{center}
%\end{figure}

\begin{table}[t]
\caption{Performance with VoIP traffic and $12$ orthogonal channels available} {
\begin{center}
\footnotesize
\begin{tabular}{|c|c|c|c|}
  \hline
  \textbf{Number} & \textbf{Optimal } & \textbf{End-to-end delay} & \textbf{Average } \\
  \textbf{of VoIP} & \textbf{end-to-end} & \textbf{achieved by} & \textbf{ARACHNE}\\
  \textbf{sessions} & \textbf{delay (s)} & \textbf{ARACHNE (s)} & \textbf{iterations}\\
  \hline\hline
  $4$ & $0.0001$ & $0.0002$ & $2$  \\
  \hline
  $8$ & $0.0008$ & $0.001$ & $4$  \\
  \hline
  $12$ & $0.001$ & $0.004$ & $7$  \\
  \hline
  $16$ & $0.006$ & $0.009$ & $11$  \\
  \hline
  $20$ & $0.011$ & $0.016$ & $14$  \\
  \hline
  $24$ & $0.014$ & $0.019$ & $18$  \\
\hline
\end{tabular}
\end {center}}\label{tableVoIP11}
\end{table}

\begin{table}
\caption{Performance with VoIP traffic and $3$ orthogonal channels available} {
\begin{center}
\footnotesize
\begin{tabular}{|c|c|c|c|}
  \hline
 \textbf{Number} & \textbf{Optimal } & \textbf{End-to-end delay} & \textbf{Average } \\
  \textbf{of VoIP} & \textbf{end-to-end} & \textbf{achieved by} & \textbf{ARACHNE}\\
  \textbf{sessions} & \textbf{delay (s)} & \textbf{ARACHNE (s)} & \textbf{iterations}\\
  \hline\hline
  $4$ & $0.0009$ & $0.0012$ & $2$  \\
  \hline
  $8$ & $0.0014$ & $0.0028$ & $3$  \\
  \hline
  $12$ & $0.0044$ & $0.0069$ & $4$  \\
  \hline
  $16$ & $0.0092$ & $0.014$ & $8$  \\
  \hline
  $20$ & $0.016$ & $0.02$ & $9$  \\
  \hline
  $24$ & $0.019$ & $0.024$ & $12$  \\
\hline
\end{tabular}
\end {center}}\label{tableVoIP3}
\end{table}

\subsection{Simulations with saturated UDP traffic and VoIP}
%We present our simulation experiments and the interpretations thereof, in what follows.
%{\bf ***** SAY HOW MANY INTERFACES ARE USED PER NODE ******}
%\textbf{\em Varying the number of clients.}
To begin with, we opt to observe the variation in the total network throughput, %while
as we increase the number of source-destination pairs.
For this, we progressively increase the number of associated clients from 5 to 70.
Here the network consists of 10 mesh APs, each of which is equipped with 2 wireless interfaces for the backhaul communication.
%We compare ARACHNE against \emph{Hyacinth}, the single-channel strategy (the same channel is used for all interfaces in the network) and a random channel allocation scheme.
%Fig. \ref{cbr_sta} depicts the throughput results.
We observe in Fig. \ref{cbr_sta} that the performance with ARACHNE is similar to the other 3 policies, when the number of clients (and therefore the communication load) is low. However, with increased load ARACHNE manages to provide much higher end-to-end throughputs (up to {\bf 85}\% difference!), due to its efficient end-to-end channel selection strategy. Hyacinth is designed especially for wireless internet traffic which is directed to/from the wireless gateways. In our simulation scenario where saturated UDP traffic is supported, the tree-based architecture is incapable to support dynamic traffic variations in the network (especially in high communication load). In low load conditions Hyacinth performs close to ARACHNE. Contrarily, when the number of clients in the networks increases the performance drops and the throughput saturation point is reached quite quickly.

\begin{figure}[t]
\centering
\subfigure[Average end-to-end delay.]{
\includegraphics[width=2.5in]{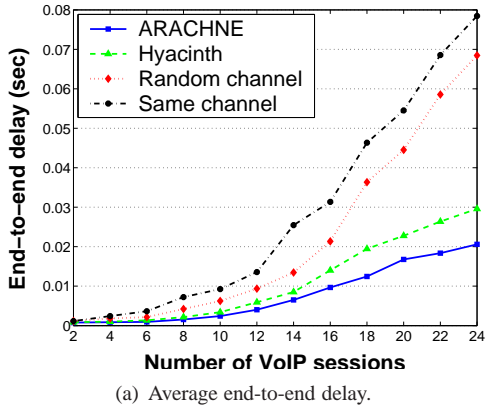}\label{voip_delay}}\\
\subfigure[Average dropped data.]{
\includegraphics[width=2.5in]{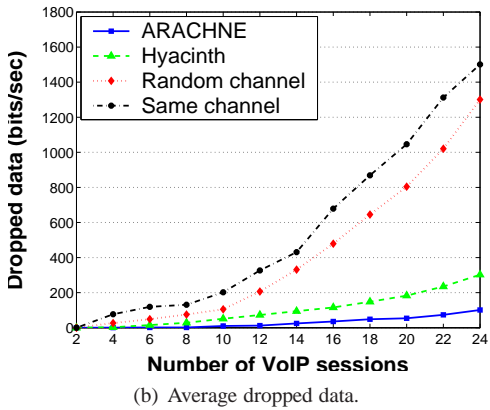}\label{voip_dropped}}
\caption{VoIP simulation results.}
\label{fig:voip} %\vspace{-0.3in}
\end{figure}

Next,
we examine how close ARACNE is performing to optimal and its' scalability. We measure the total end-to-end network throughput
%while the number of the mesh APs in the networks is increased
as we increase the number of APs (i.e., the mesh density)
from 10 to 80 (we increase the number of clients along with the mesh APs:
10 APs - 40 clients, 20 APs - 80 clients, etc.).
Note that the interference is dynamically changing while the number of the APs in the network increases;
the airtime cost metric manages to effectively capture the varying
co-channel interference.

The first objective in our study is to examine how efficient ARACHNE approaches the optimal network performance. Our methodology includes the incorporation of the operational parameters in the IBM ILOG CPLEX solver while the number of the existing APs in the network increases. The optimal channel allocation then is applied in our simulation scenarios (OPNET). We apply ARACHNE under the same operational parameters of the network and compare the achieved throughput to the optimal throughput. In table \ref{tableUDP11}, we observe the total network throughput that is achieved by ARACHNE after a small number of iterations till convergence, compared to the optimal network throughput (when 12 orthogonal channels are available). Moreover, table \ref{tableUDP3} contains similar results when there are just 3 available orthogonal channels availabe to allocate in the network. A general outcome is that ARACHNE is able to approach the optimal network performance and scales efficiently while the network topology and the load vary. The best performance is achieved within the case of 12 orthogonal channels, where ARACHNE gets as input more frequencies to allocate and converges to channel allocation solutions close to optimal (by applying its' sophisticated policy).

\begin{figure}[t]
\centering
\subfigure[Dartmouth traces.]{
\includegraphics[width=2.5in]{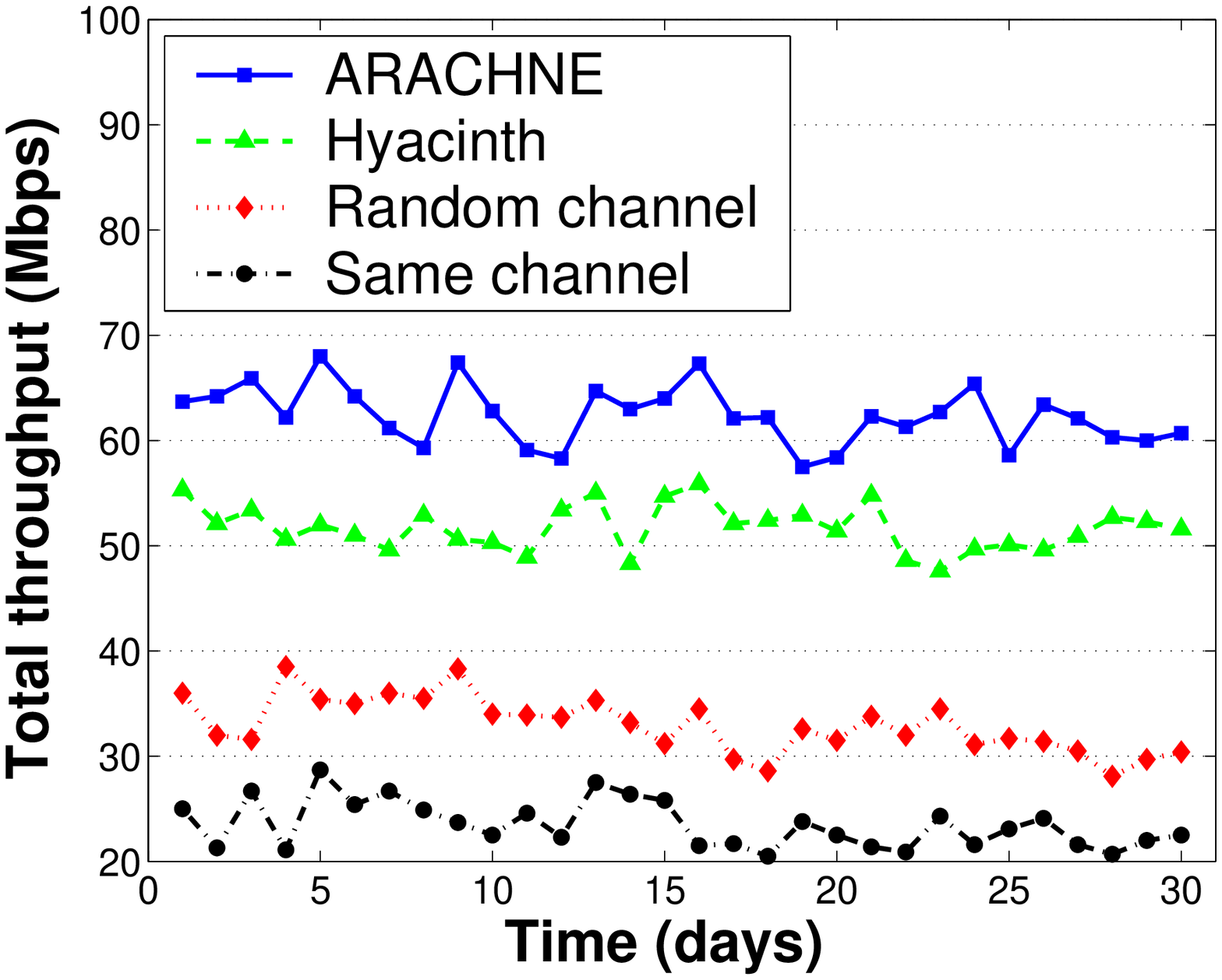}\label{real2dart}}\\
\subfigure[IBM traces.]{
\includegraphics[width=2.5in]{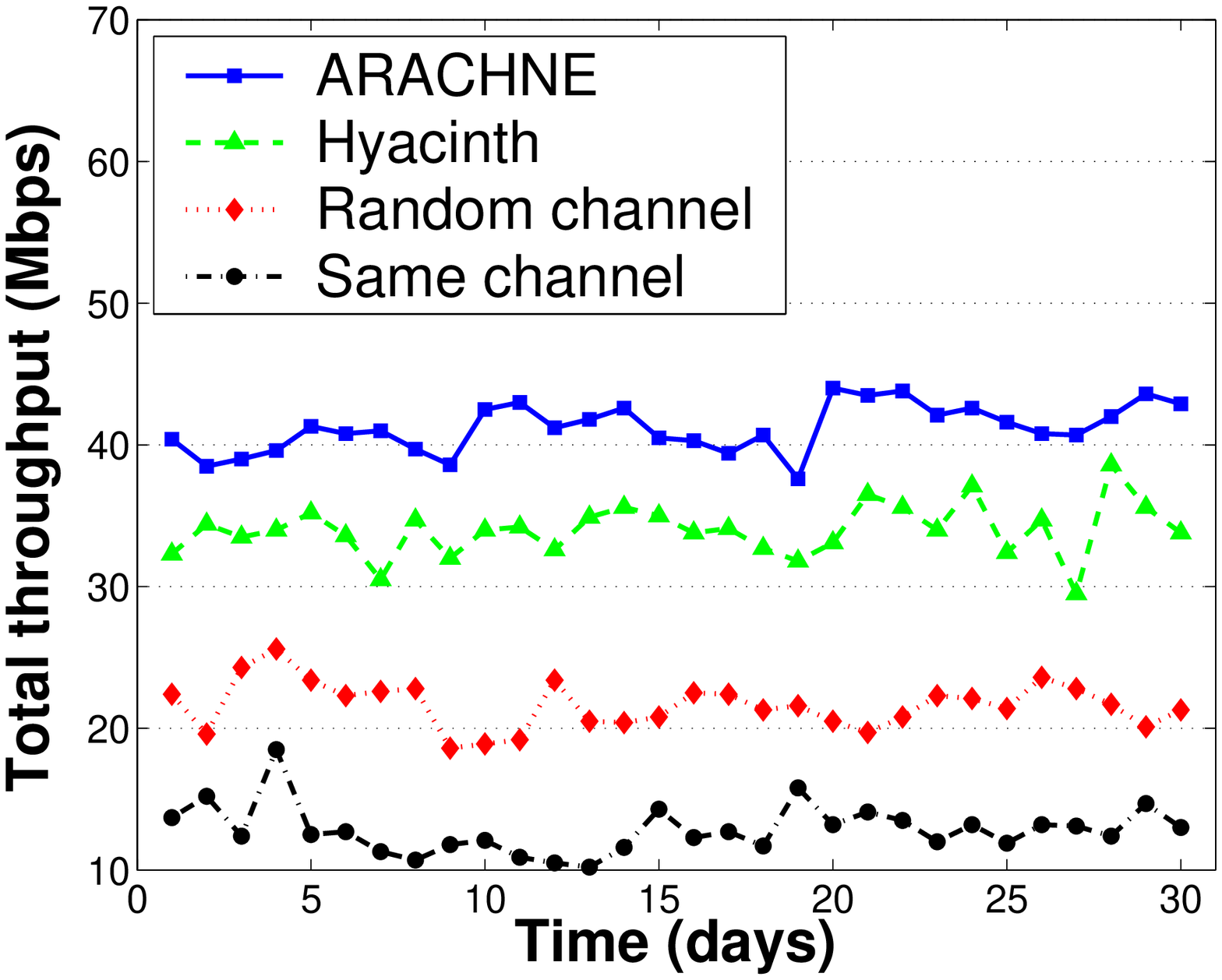}\label{real2ibm}}
\caption{Simulation results with real traces: 2 interfaces per node.}
\label{fig:real1} %\vspace{-0.2in}
\end{figure}

\begin{figure}[t]
\centering
\subfigure[Dartmouth traces.]{
\includegraphics[width=2.5in]{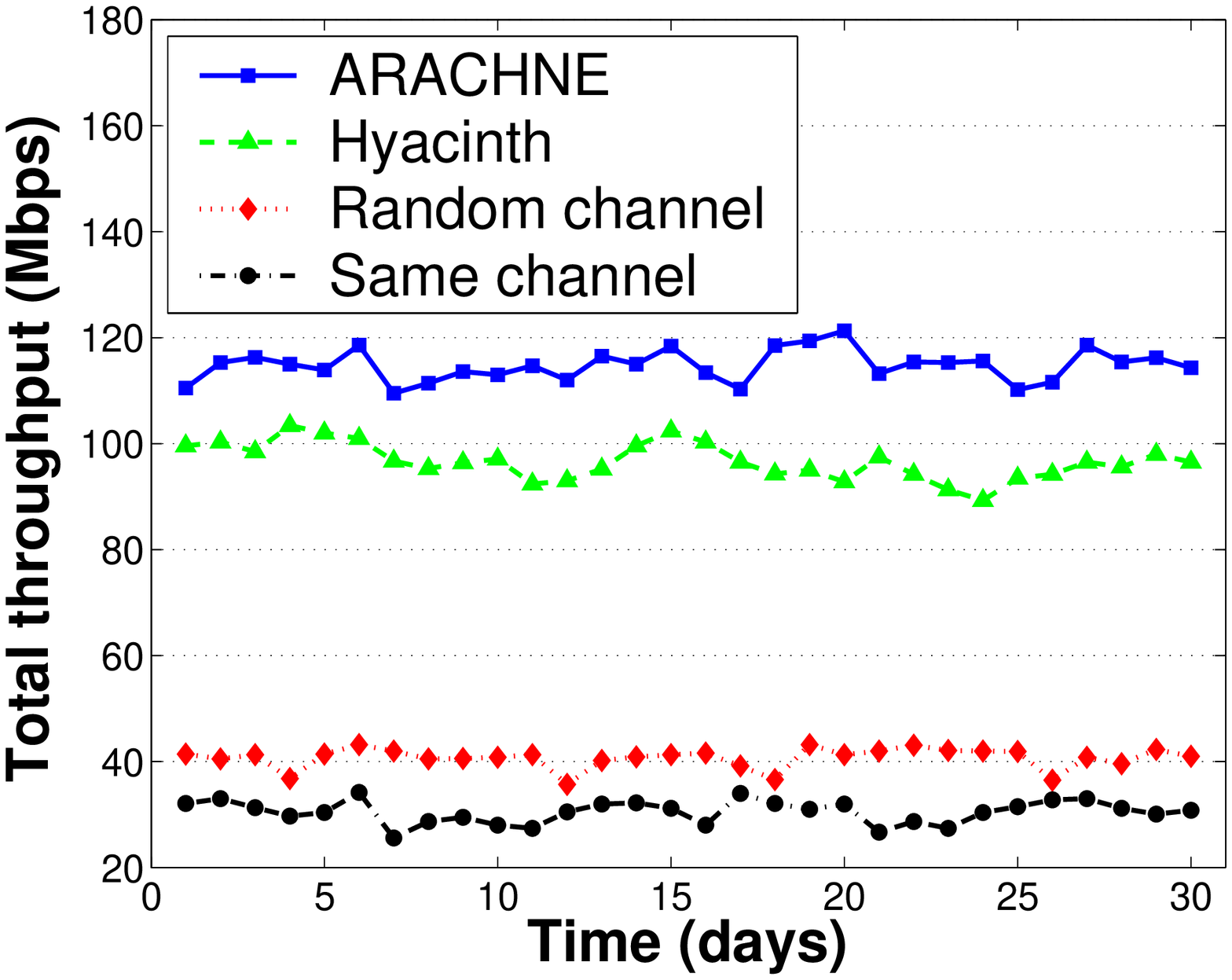}\label{real4dart}}\\
\subfigure[IBM traces.]{
\includegraphics[width=2.5in]{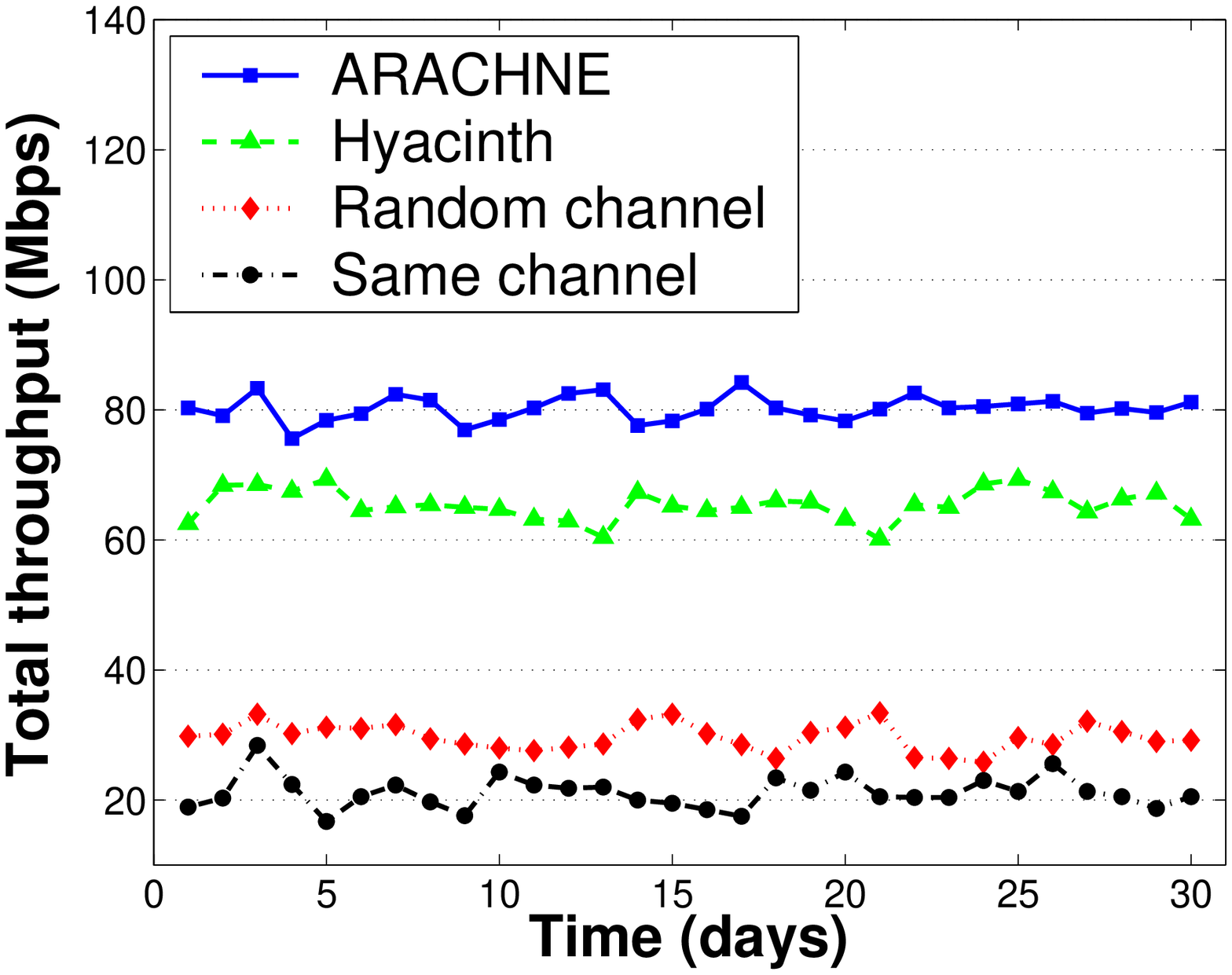}\label{real4ibm}}
\caption{Simulation results with real traces: 4 interfaces per node.}
\label{fig:real2} %\vspace{-0.2in}
\end{figure}

Fig. \ref{cbr_ap} depicts the network performance in terms of total network throughput.
%It is clear
We observe that ARACHNE is able to adapt to  %our protocol adapt its' behavior to the
topology and %the
load variations, and therefore it manages to % optimizes
provide the most beneficial
%the
 selection of the available channels at the mesh backhaul. The tree-based architecture that is introduced by Hyacinth does not scale well and therefore the achieved performance is worst.
% effect and provide a sophisticated channel allocation.
%Contrarily, the other two approaches do not take into account the interference and so they are not able to provide efficient network operation (especially in high load network operation).

%\textbf{\em Simulating ARACHNE with VoIP traffic.}
In order to observe the performance of our protocol with delay-sensitive data exchange, we utilize varying, parallel, end-to-end VoIP traffic sessions.
%Fig. \ref{voip_delay} and \ref{voip_dropped} depict the network performance with VoIP.
%when we support VoIP, which is a QoS sensitive application, in our wireless mesh network. The simulation results are performed by varying the number of the  VoIP sessions that are supported in parallel. In
Tables \ref{tableVoIP11} and \ref{tableVoIP3} present the average end-to-end delay that ARACHNE achieves (when 3 and 12 orthogonal channels are supported in the network). The optimal end-to-end delay values are used (again) as a benchmark. We observe that ARACHNE can perform close to optimal after a lightweight execution in the network (i.e. converges after a small number of iterations).

Fig. \ref{voip_delay} depicts the average end-to-end delay of VoIP packet transmissions. % from the sender to the receiver.
We observe that ARACHNE achieves low %er
end-to-end delays, since it considers the load of the individual links across a route in the process of assigning frequencies to wireless interfaces.
Recall here that ARACHNE is expected to provide high benefits when operating in conjunction with load-aware routing protocols, such as RM-AODV (this is case with our simulations), since they both take into consideration the load that is experienced by individual links.
%However, ARACHNE's operations is independent of the routing protocol that is employed, as we discuss in section \ref{sec:scope}.
Furthermore,
%compared to same and random channel selection approaches. In particular, our protocol provides efficient access level and mesh backhaul communication and combined with the RM-AODV protocol (which uses the airtime metric in the routing process, achieving in this way low routing end-to-end delays) can effectively support several VoIP sessions in parallel.

Fig. \ref{voip_dropped} shows the average number of  dropped data packets %in the VoIP communication
due to channel errors and contention.
The performance of ARACHNE is impressive, since  %while the data
packet dropping is kept in very low levels, as compared to Hyacinth and to the other strategies. Hyacinth does not consider the channel allocation at the access level and therefore, it can support less VoIP sessions than ARACHNE can (which provides end-to-end channel selection and therefore the provided QoS to the VoIP clients is higher).

%The data dropping in the other two approaches is extremely high (especially in high load conditions) and therefore the VoIP application cannot be supported.
%
%\begin{figure*}[th]
%\centering
%\subfigure[Dartmouth traces (2 interfaces).]{
%\includegraphics[width=1.5in]{figures/throughput2_dart.eps}\label{real2dart}}
%\subfigure[IBM traces (2 interfaces).]{
%\includegraphics[width=1.5in]{figures/throughput2_ibm.eps}\label{real2ibm}}
%\subfigure[Dartmouth traces (4 interfaces).]{
%\includegraphics[width=1.5in]{figures/throughput4_dart.eps}\label{real4dart}}
%\subfigure[IBM traces (4 interfaces).]{
%\includegraphics[width=1.5in]{figures/throughput4_ibm.eps}\label{real4ibm}}
%\caption{\small Simulation results with real traces: ARACHNE is predominant!}
%\label{fig:real} %\vspace{-0.2in}
%\end{figure*}
%\subsection{Real traces}

\begin{figure}[t]
\centering
\subfigure[Hyacinth performance.]{
\includegraphics[width=2.5in]{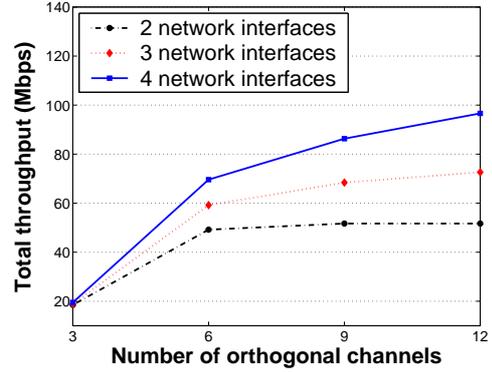}\label{real2dart1}}\\
\subfigure[ARACHNE performance.]{
\includegraphics[width=2.5in]{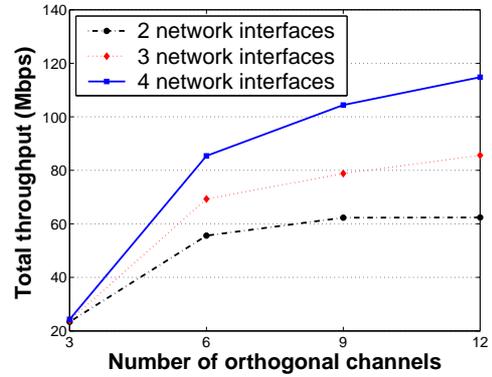}\label{real2ibm1}}
\caption{Simulation results with real traces: Varying the number or the supported interfaces and the available channels in the network with Dartmouth traces.}
\label{fig:real3} %\vspace{-0.2in}
\end{figure}

\begin{figure}[t]
\centering
\subfigure[Hyacinth performance.]{
\includegraphics[width=2.5in]{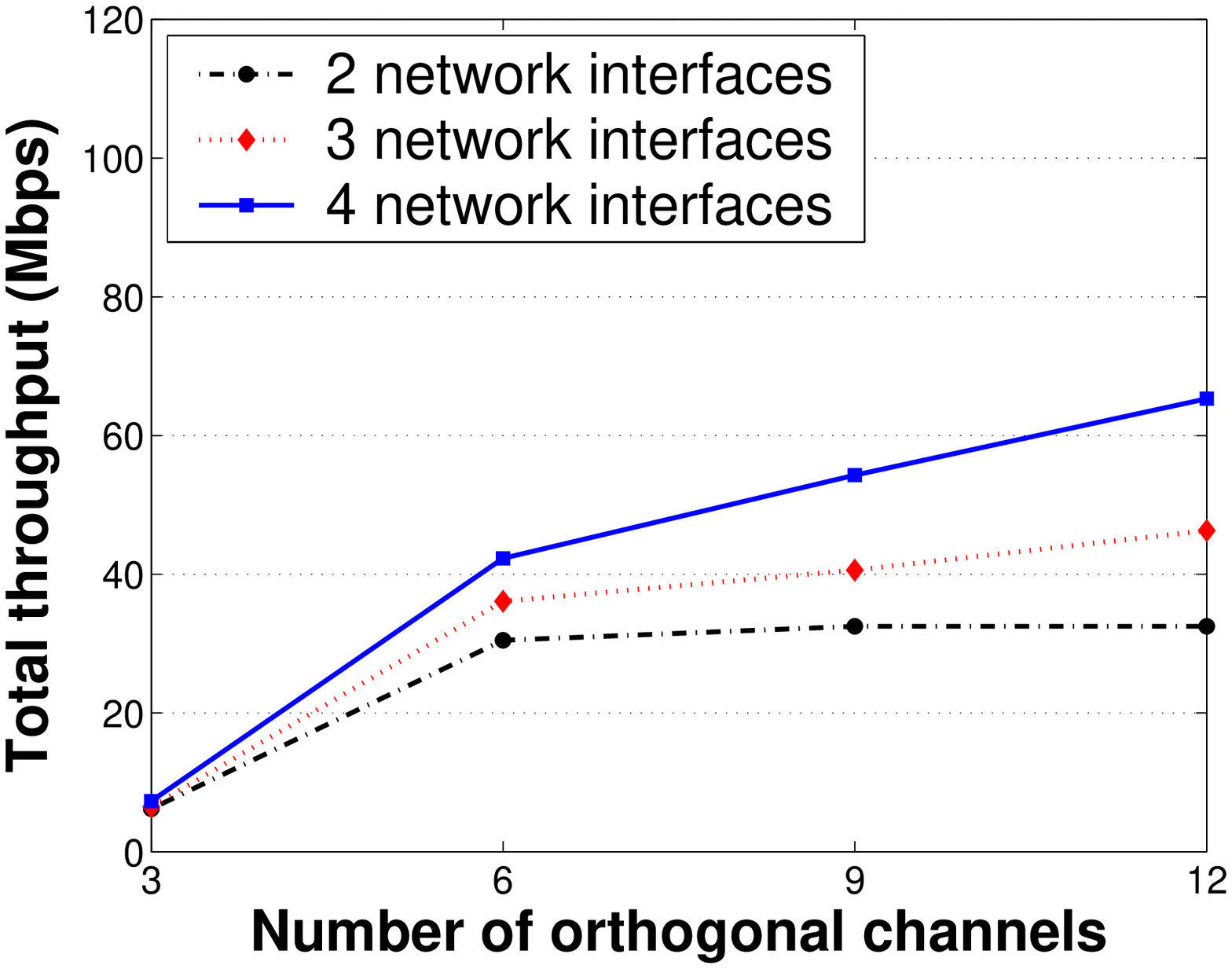}\label{real4dart1}}\\
\subfigure[ARACHNE performance.]{
\includegraphics[width=2.5in]{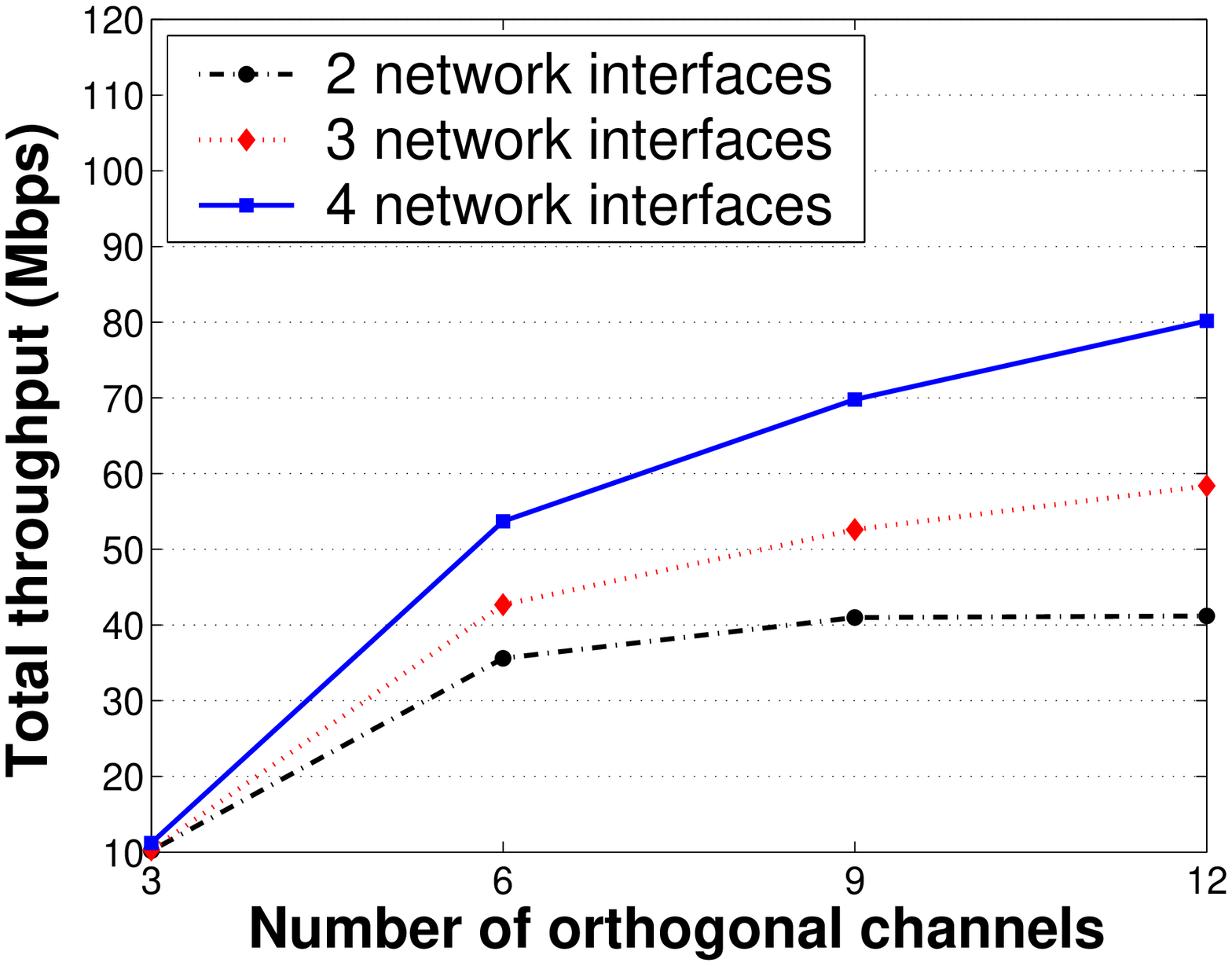}\label{real4ibm1}}
\caption{Simulation results with real traces: Varying the number or the supported interfaces and the available channels in the network with IBM traces.}
\label{fig:real4} %\vspace{-0.2in}
\end{figure} 

\subsection{Simulation results with real traces}
Next, we present our simulation experiments with real traces from Dartmouth College and IBM.
This will reveal the behavior of ARACHNE in more realistic scenarios.
By analyzing the SNMP logs from each AP, we derive  %get
the dynamic behavior of the aggregated traffic demand.
Note that these traces have been gathered from real wireless networks  %(Wireless Local Area Network)
settings.
Therefore, they represent actual traffic demands and behaviors of the APs and clients in a real network.
%%  IS YOUR FOLLOWING ARGUMENT IS NOT SUPPORTED SOMEHOW? IF WE HAVEN'T REFERENCES FOR THIS, WE CANNOT WRITE IT...
%the APs is such deployments serve a similar role and thus exhibit similar behavior as the local APs of a wireless mesh network.
%The second part of our evaluation methodology contains the use of real traces in the direction of protocol evaluation in realistic environment. We use traces from two different locations in order to strengthen our protocol evaluation and study its' performance in different operational environments. By analyzing the snmp log from each AP, we get the dynamic behavior of the aggregated traffic demand. The APs of a WLAN (Wireless Local Area Network) serve a similar role and thus exhibit similar behavior as the local APs of a wireless mesh network.
%\textbf{\em Handling the network traces in our simulations.}
The traces from Dartmouth College \cite{dartmouth} (3/2001) involve the buildings labeled as {\em AcadBldg10, SocBldg4}, and include approximately 40 APs (the traces contain the AP location; we have placed them accordingly in our simulation space).
In addition, the traces from IBM  \cite{ibm} (8/2002) have been gathered from the buildings labeled as {\em MBldg, and SBldg}.
The number of the APs that are placed in these buildings varies (the traces don't contain their locations and therefore we have simulated various random topologies with them).
Note that the number of active clients (end-hosts) varies in both trace sets and they are uniformly distributed in our experiments (the traces do not contain information about client locations).
In order to import the traces in our simulation we  periodically compute the average rate for each client, once every 3 minutes, %.  Furthermore, %Then
and we correspond the computed average values to the traffic demands of each client.

%Firstly, we use WLAN traces from Dartmouth College [?] (3/2001). In our experiments we use the traces from buildings labeled as AcadBldg10, SocBldg4 in Dartmouth College. In these buildings there are approximately 40 APs (the traces contain their location and so we have placed them accordingly in our simulation). In addition, we use WLAN traces from IBM [?] (8/2002) and particularly from the buildings labeled as MBldg, and SBldg. The number of the APs that are placed in these buildings varies (the traces don't contain their locations and therefore we have randomly placed them). We must mention that the number of the STAs that are active varies in both trace sets and they are uniformly distributed in our experiments (the traces don't contain information about the STAs locations). In order to import the traces in our simulation we compute periodically the average rate for each STA (every 3 min). Then we use this average values as the traffic demand of each STA in the network (this traffic will be routed by RM-AODV at the mesh backhaul).
%\textbf{\em ARACHNE outperforms other schemes in terms of overall throughput.}

To begin with, we consider 2 wireless interfaces at each AP for the backhaul communication.
Fig. \ref{real2dart} and \ref{real2ibm} show that ARACHNE achieves the
highest total network throughput in all cases.
Our simulation experiments reveal that channel selection at the access level plays an important role in the overall end-to-end performance.
Being based on the airtime cost metric, ARACHNE takes into account the channel conditions as well as the communication load in both access level and the mesh backhaul (not considered by the other approaches, like Hyacinth) it manages to boost the network throughput.
Furthermore, Fig. \ref{real4dart} and \ref{real4ibm} depict the performance of the compared strategies when the mesh APs are equipped with 4 wireless interfaces.
The results are similar as above.

Last but not least, we perform some simulations where we vary both the number of available channels and the number of the wireless interface supported by each AP. Fig. \ref{real2dart1}, \ref{real2ibm1}, \ref{real4dart1} and \ref{real4ibm1} show that ARACHNE achieves higher throughput values compared to Hyacinth \cite{hyacinth}. Throughput improvement is significantly increasing while we the number of available channels and the wireless interfaces increase.

\section{Conclusions and Future Work}
\label{sec:conclusion}
In this paper we presented a channel allocation methodology with main objective to maximize the end-to-end throughput in wireless multi-radio mesh networks. The modeling of the channel allocation problem inspired the design of a routing-aware channel selection mechanism (ARACHNE). ARACHNE adopts a metric that provides an estimation of the average packet transmission delay in a communication channel. In addition, ARACHNE captures the interference effects in the networks and applies an end-to-end channel selection policy that manages to provide approximately the maximum end-to-end network throughput. Our work was compared against 3 other representative channel selection approaches and we showed that it outperforms all of them, for different traffic scenarios and network densities. Moreover, the simulation results showed that ARACHNE converges after a small number of iterations, proving in this way its' lightweight operation.

%\section{Acknowledgement}
%\label{sec:ack}
%The authors acknowledge the support of the Greek Secretariat
%of Research and Technology through a PENED 2003
%grant and the support of the European Commission OPNEX STREP (FP7-224218).

%\clearpage
\bibliographystyle{unsrt}	
\bibliography{main}

\begin{thebibliography}{10}

\bibitem{microsoftmesh}
Microsoft Research.
\newblock Self-organizing neighborhood wireless mesh networks project.
\newblock http://research.microsoft.com/mesh/.

\bibitem{meraki}
Meraki~Networks Inc.
\newblock http://meraki.net.

\bibitem{Athanasiou}
G.~Athanasiou, T.~Korakis, O.~Ercetin, and L.~Tassiulas.
\newblock {Dynamic Cross-Layer Association in 802.11-based Mesh Networks}.
\newblock In {\em IEEE INFOCOM}, 2007.

\bibitem{gibbsInfocom}
B.~Kauffmann, F.~Baccelli, A.~Chainteau, V.~Mhatre, K.~Papagiannaki, and
  C.~Diot.
\newblock {{Measurement-Based Self Organization of Interfering 802.11 Wireless
  Access Networks}}.
\newblock In {\em IEEE INFOCOM}, 2007.

\bibitem{rozner}
E.~Rozner, Y.~Mehta, A.~Akella, and L.~Qiu.
\newblock {{Trafic-Aware Channel Assignment in Enterprise Wireless LANs}}.
\newblock In {\em IEEE ICNP}, 2007.

\bibitem{maxchop}
A.~Mishra, V.~Shrivastava, D.~Agarwal, and S.~Banerjee.
\newblock {{Distributed Channel Management in Uncoordinated Wireless
  Environments}}.
\newblock In {\em ACM MOBICOM}, 2006.

\bibitem{hyacinth}
A.~Raniwala and T.~Chiueh.
\newblock {Architecture and Algorithms for an IEEE 802.11-Based Multi-Channel
  Wireless Mesh Network}.
\newblock In {\em IEEE INFOCOM}, 2005.

\bibitem{mobicom05}
M.~Alicherry, R.~Bhatia, and L.~Li.
\newblock {Joint Channel Assignment and Routing for Throughput Optimization in
  Multi-radio Wireless Mesh Networks}.
\newblock In {\em ACM MOBICOM}, 2005.

\bibitem{80211s}
{IEEE 802.11s: Wireless LAN Medium Access Control (MAC) and Physical Layer
  (PHY) Specifications: Simple Efficient Extensible Mesh (SEE-Mesh) Proposal}.

\bibitem{opnet}
{OPNET-Radio/Wireless Models}.
\newblock http://www.opnet.com.

\bibitem{LCCS}
J.~Geier.
\newblock {{Assigning 802.11b Access Point Channels}}.
\newblock In {\em WiFi planet}, 2002.

\bibitem{sumanb}
A.~Mishra, V.~Brik, S.~Banerjee, A.~Srinivasan, and W.~Arbaugh.
\newblock {{A Client-Driven Approach for Channel Management in Wireless
  {LAN}s}}.
\newblock In {\em IEEE INFOCOM}, 2006.

\bibitem{clifford}
D.~J. Leith and P.~Clifford.
\newblock {A Self-Managed Distributed Channel Selection Algorithm for WLANs}.
\newblock In {\em WiOPT}, April 2006.

\bibitem{rozner1}
P.H. Pathak.
\newblock {{A Survey of Network Design Problems and Joint Design Approaches in
  Wireless Mesh Networks}}.
\newblock In {\em IEEE Communications Surveys and Tutorials}, 2011.

\bibitem{bianchi}
G.~Bianchi.
\newblock {Performance Analysis of the 802.11 DCF}.
\newblock In {\em IEEE JSAC, vol.18, pp.535-547}, March 2000.

\bibitem{gupta}
N.~Gupta and P.~R. Kumar.
\newblock {A Performance Analysis of the 802.11 Wireless LAN Medium Access
  Control}.
\newblock In {\em Comm. in Inform. and Systems, Vol.3, No.4, pp. 279-304},
  September 2004.

\bibitem{hidden}
T-C Hou, L-F Tsao, and H-C liu.
\newblock {Analyzing the Throughput of 802.11 DCF Scheme With Hidden Nodes}.
\newblock In {\em IEEE VTC}, 2003.

\bibitem{kumar}
A.~Kumar, E.~Altman, D.~Miorandi, and M.~Goyal.
\newblock {New Insights from a Fixed Point Analysis of Single Cell IEEE 802.11
  WLANs"}.
\newblock In {\em IEEE INFOCOM}, 2005.

\bibitem{anomaly}
M.~Heusse, F.~Rousseau, G.~Berger-Sabbatel, and A.~Duda.
\newblock {Performance Anomaly of 802.11b}.
\newblock In {\em IEEE INFOCOM}, 2003.

\bibitem{bertsekas}
D.~Bertsekas and R.~Gallager.
\newblock {Data Networks}.
\newblock In {\em Englewood Cliffs, NJ: Prentice-Hall}, 1992.

\bibitem{niculescu}
D.~Niculescu.
\newblock {Interference Map for 802.11 Networks}.
\newblock In {\em ACM IMC}, 2007.

\bibitem{niculescu1}
D.~Niculescu, S.~Bhatnagar, and S.~Ganguly.
\newblock {Channel Assignment for Wireless Meshes with Tree Topology}.
\newblock In {\em COMM}, 2010.

\bibitem{mdg}
I.~Broustis, K.~Papagiannaki, S.~V. Krishnamurthy, M.~Faloutsos, and V.~Mhatre.
\newblock {{MDG: Measurement-Driven Guidelines for 802.11 WLAN Design}}.
\newblock In {\em ACM MOBICOM}, 2007.

\bibitem{mdg1}
I.~Broustis, K.~Papagiannaki, S.~V. Krishnamurthy, M.~Faloutsos, and V.~Mhatre.
\newblock {{Measurement-Driven Guidelines for 802.11 WLAN Design}}.
\newblock In {\em IEEE/ACM Transactions on Networking}, June 2010.

\bibitem{infocom08}
L.~Dai, Y.~Xue, B.~Chang, Y.~Cao, and Y.~Cui.
\newblock {Integrating Traffic Estimation and Routing Optimization for
  Multi-Radio Multi-Channel Wireless Mesh Networks}.
\newblock In {\em IEEE INFOCOM}, 2008.

\bibitem{ilog}
{IBM ILOG CPLEX Optimizer}.
\newblock
  http://www-01.ibm.com/software/integration/optimization/cplex-optimizer/.

\bibitem{dartmouth}
{Dartmouth Campus-Wide Wireless Traces}.
\newblock http://crawdad.cs. dartmouth.edu/meta.php?name=dartmouth/campus.

\bibitem{ibm}
{IBM Wireless Traces}.
\newblock http://nms.lcs.mit.edu/~mbalazin/wireless/.

\end{thebibliography}


%% References


%\bibliographystyle{elsarticle-num}






% \end{thebibliography}

\end{document}